\documentclass[10pt, letterpaper]{article}

\usepackage[a4paper, total={6.5in, 8in}]{geometry}
\usepackage{graphicx}
\usepackage{xcolor}
\usepackage{ragged2e}
\usepackage{placeins}

\newcommand{\beginsupplement}{%
        \setcounter{table}{0}
        \renewcommand{\thetable}{S\arabic{table}}%
        \setcounter{figure}{0}
        \renewcommand{\thefigure}{S\arabic{figure}}%
     }

\begin{document}

\title{Cognitive chimera states in human brain networks}
\maketitle

\begin{large}
\noindent Kanika Bansal\textsuperscript{1,2,3,*},
Javier O. Garcia\textsuperscript{1,4},
Steven H. Tompson\textsuperscript{1,4},
Timothy Verstynen\textsuperscript{5},
Jean M. Vettel\textsuperscript{1,4,6},
Sarah F. Muldoon\textsuperscript{3,7,*}\\
\end{large}

\noindent\textsuperscript{1} Human Sciences, US Army Research Laboratory, Aberdeen Proving Ground, MD 21005, USA\\
\textsuperscript{2} Department of Biomedical Engineering, Columbia University, New York, NY 10027, USA\\
\textsuperscript{3} Mathematics Department, University at Buffalo -- SUNY, Buffalo, NY 14260, USA\\
\textsuperscript{4} Department of Biomedical Engineering, University of Pennsylvania, Philadelphia, PA 19104, USA\\
\textsuperscript{5} Department of Psychology, Carnegie Mellon University, Pittsburgh, PA 15213, USA\\
\textsuperscript{6} Department of Psychological and Brain Sciences, University of California, Santa Barbara, CA 93106 , USA\\
\textsuperscript{7} CDSE Program and Neuroscience Program, University at Buffalo -- SUNY, Buffalo, NY 14260, USA\\

\noindent \textsuperscript{*}Correspondence should be addressed to KB (phy.kanika@gmail.com) and SFM (smuldoon@buffalo.edu).

\newpage
\begin{abstract}
\noindent The human brain is a complex dynamical system that gives rise to cognition through spatiotemporal patterns of coherent and incoherent activity between brain regions. As different regions dynamically interact to perform cognitive tasks, variable patterns of partial synchrony can be observed, forming chimera states. We propose that the emergence of such states plays a fundamental role in the cognitive organization of the brain, and present a novel cognitively-informed, chimera-based framework to explore how large-scale brain architecture affects brain dynamics and function. Using personalized brain network models, we systematically study how regional brain stimulation produces different patterns of synchronization across predefined cognitive systems. We then analyze these emergent patterns within our novel framework to understand the impact of subject-specific and region-specific structural variability on brain dynamics. Our results suggest a classification of cognitive systems into four groups with differing levels of subject and regional variability that reflect their different functional roles.

\end{abstract}

\newpage

\section*{Introduction}

\noindent Rhythmic behavior is ubiquitous in complex systems, and a diverse set of research has examined how interacting system elements come together to form synchronized, coherent behavior across domains that span biological \cite{Winfree1967,Glass2001}, social \cite{Jia2017}, and engineered \cite{Strogatz2005} settings. However, the emergence of complete system-wide synchronization might not always provide the best description of system dynamics. In many systems, states of partial synchrony have been observed, where a system organizes in separate domains of synchronized elements \cite{Bertolero2015,Motter2010,Pecora2014}. This is particularly true in the human brain, where patterns of neurophysiological activity evolve rapidly, showing transient domains of synchronization across subsets of brain regions \cite{Gray1994,Lachaux1999,Bertolero2015}. In the last decade, the rise of network neuroscience approaches \cite{Bullmore2009,Bassett2017} have demonstrated a foundational role for partial synchrony among separate cognitive sub-networks, where the underlying architecture of the brain ensures efficient integration of sensory input with stored knowledge while also segregating task-irrelevant information to support cognition \cite{Shine2016,Vatansever2017}. However, the fundamental principles and constraints that subserve the intricate timing and specificity of these time-evolving patterns of synchrony are not well understood \cite{Deco2015}.

%The interplay of synchrony among a set of cognitive systems in the brain has been shown to underlie sleep-wake states \cite{Killgore2010}, variability in task performance \cite{Shine2016}, and the continuum between healthy and disease states \cite{Menon2011,Fornito2015}. 

The dynamical systems framework of \textit{chimera states} offers a powerful tool to study the evolution of coherent and incoherent dynamics in oscillating systems such as the brain. A chimera state emerges when a system of oscillators evolves into two subsets of mutually coherent and incoherent populations \cite{kuramoto2002coexistence,abrams2004chimera}. Although chimera states represent a natural link between coherent and incoherent dynamics \cite{Motter2010,Omelchenko2008,LAING2009}, they were initially only explored analytically and their relationship to physical systems was unknown.  In fact, it wasn't until almost a decade after their theoretical discovery that chimera states were demonstrated experimentally in opto-electronic \cite{hagerstrom2012experimental}, and chemical \cite{Tinsley2012} oscillators, finally establishing their connection with real-world systems. Recently, the chimera framework has been extended to include states of multi-chimera \cite{Omelchenko2013,Vuellings2014}, traveling chimera \cite{Jianbo2014}, amplitude chimera, and chimera death \cite{Zakharova2016}, demonstrating the versatility of the framework to describe critical system dynamics.

%Chimera states were initially only explored in mathematical models as a natural link between coherent and incoherent dynamics \cite{Motter2010,Omelchenko2008,LAING2009}. Almost a decade after their theoretical discovery, chimera states were demonstrated experimentally in opto-electronic \cite{hagerstrom2012experimental} and chemical \cite{Tinsley2012} oscillators, confirming their connection with real-world systems \cite{Andrzejak2016,Tognoli2014,Motter2013}. Consequently, chimera states have become a rapidly growing field of research \cite{panaggio2015chimera,Kemeth2016}, both analytically \cite{Sethia2014,Schmidt2014,Ashwin2015,Schoell2016} and experimentally \cite{Martens2013,Larger2013,Rosin2014,Viktorov2014,Larger2015,Hart2016}. Recently, the chimera framework has been extended to include states of multi-chimera \cite{Omelchenko2013,Vuellings2014}, traveling chimera \cite{Jianbo2014}, amplitude chimera, and chimera death \cite{Zakharova2016}. \KBJV{Collectively, these results have demonstrated the versatility of chimera states to encapsulate critical components of system dynamics.}

Because of its natural ability to describe patterns of partial synchronization, the chimera framework has an intuitive utility for augmenting our understanding of the brain. Patterns of synchronization between cognitive systems are thought to form the basis of cognition, and the interplay of synchrony among subsets of brain regions has been shown to underlie sleep-wake states \cite{Killgore2010}, variability in task performance \cite{Shine2016}, and the continuum between healthy and disease states \cite{Menon2011,Fornito2015}. Indeed, recent work has speculated that similarities exist between chimera states and brain dynamics during unihemispheric sleep \cite{panaggio2015chimera}, the transition to a seizure state in epilepsy \cite{Andrzejak2016}, and coordinated finger tapping exercises \cite{Tognoli2014}. Fundamentally, these dynamics are the result of complex interactions between neuronal populations and are often modeled using networks of coupled oscillators \cite{Ashwin2016}. As a result, despite the intuitive similarities between chimera states and brain dynamics, much of the work relating chimera dynamics to neuroscience thus far has focused on understanding chimera states at the level of neuronal networks, using mathematical modeling of networks of individual neurons with fewer elements and/or simplified connection topologies \cite{Omelchenko2013,Vuellings2014,Hizanidis2014,Bera2016b}. Only recently have neuronal models been used to examine the possibility of the emergence of chimera-like states within large-scale brain networks derived from two well-characterized animal brains -- \textit{C. elegans} \cite{Hizanidis2016} and the cat cortex \cite{SANTOS2017}. However, even in these instances, the network connectivities were modified for simplicity. Thus, there remains a gap between studies of chimera states and applications to large-scale functional patterns of brain activity thought to underlie cognition. This largely reflects the computational complexity of modeling whole-brain dynamics and identifying an informative, yet simplified, model of cognitive processing.

%Recently, studies have confirmed the emergence of chimera states in models describing the dynamics of coupled neurons. Using simplified connectivity schemes of two well-characterized animal brains -- \textit{C. elegans} and cat, researchers have demonstrated the role of both chemical and electrical signaling in the emergence of chimera states \cite{Hizanidis2016,SANTOS2017}. A relatively larger portion of research has explored the role of local and non-local coupling schemes in the appearance of chimera states \cite{{Omelchenko2013},{Vuellings2014},{Hizanidis2014},{Bera2016b}}. While these approaches studying small neuronal networks have confirmed the emergence of chimera states under a variety of coupling conditions, they are vastly restricted to networks with simplified topologies and dynamics when compared to the complexity of the human brain. Thus, research has not yet examined how chimera states relate to large-scale functional patterns of brain activity thought to underlie cognition. This largely reflects the computational complexity of modeling whole-brain dynamics and identifying an informative, yet simplified, model of cognitive processing. 

Here, we bridge this gap by presenting a novel cognitively-informed, chimera-based framework combined with \textit{in silico} experiments, where we leverage the existence of a core set of predefined cognitive systems that constitute the functional organization of the brain \cite{Muldoon2016,Power2011}. Consequently, our novel framework keeps the computational complexity of the analysis to a minimum while providing the unique ability to connect chimera states as an underlying basis for cognition.  Using personalized brain network models (BNMs), we study cognitive system-level patterns of synchrony that emerge across 76 brain regions within 9 cognitive systems as the result of regional brain stimulation. Our analysis focuses on how brain architecture relates to the frequency and types of dynamical patterns produced after stimulation. More specifically, we aim to answer two questions: (i) do patterns of synchronization observed for each cognitive system depend on what region was stimulated (region-specific effects), and (ii) does structural variability between participants decrease the consistency of patterns observed for each cognitive system (subject-specific effects)? 

%To investigate these questions, we build individual-specific BNMs using structural connectivity derived from neuroimaging data, and we systematically study the results of computational stimulation across 76 brain regions assigned to 9 cognitive systems using BNMs for 30 healthy individuals.

From our \textit{in silico} experiments, we observe different patterns of synchronization that can be classified into three dynamical states: (i) a coherent state of global synchrony, (ii) a chimera state with coexisting domains of synchrony and de-synchrony, and (iii) a metastable state with an absence of any large-scale stable synchrony. Our results demonstrate rich diversity in the states produced across all nine cognitive systems, including variability in patterns based on both region-specific and subject-specific structural variability. Critically, all nine cognitive systems give rise to chimera states and this likely reflects the foundational role that partial synchrony serves in large-scale brain function. Neuronal dynamics must concurrently segregate specialized processing while integrating localized functions for coordinated, cohesive cognitive performance. Our novel chimera-based framework provides an avenue to study how dynamical states give rise to variability in cognitive performance, providing the first approach that can uncover the link between chimera states and cognitive system functions that subserve human behavior.

\section*{Results}

\begin{figure}
\centering
\includegraphics [width=0.5\linewidth, keepaspectratio]
{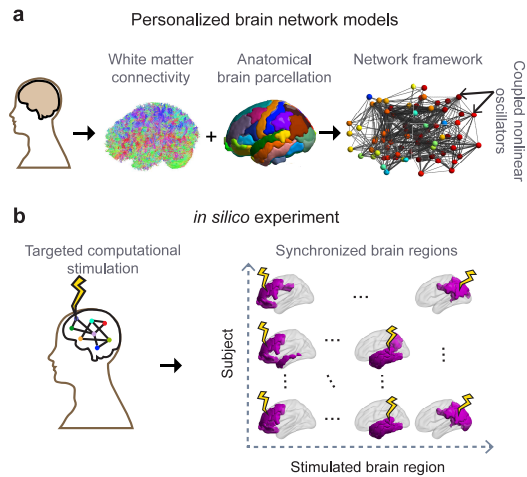}
\caption{Design of the \textit{in silico} experiments. (a) We construct personalized brain network models by estimating white matter anatomical connectivity of the brain using diffusion spectrum imaging. This connectivity is combined with a brain parcellation scheme with $76$ cortical and subcortical regions to obtain large scale connectivity map of the regional brain volume. These regions constitute the nodes of the structural brain network, whose dynamics are simulated by nonlinear Wilson-Cowan oscillators, coupled through the structural connectivity map of a given individual (see Methods). (b) In the resulting data-driven models of the spatiotemporal dynamics of the brain, each brain region is systematically stimulated across a cohort of $30$ subjects. The spread of the stimulation is measured through synchronization within the brain network.}
\label{fig:conceptual_schematic}
\end{figure}

We first build subject-specific BNMs using anatomical connectivity derived from diffusion spectrum imaging data of 30 healthy individuals \cite{Yeh2016}. As illustrated in Fig. \ref{fig:conceptual_schematic}a, we parcellate the brain into 76 regions (network nodes) and define weighted network edges based on structural connectivity between brain regions. The dynamics of each brain region are modeled using Wilson-Cowan oscillators (WCOs), a biologically inspired, nonlinear mean-field model of a small population of neurons \cite{Wilson1972}. The coupling between regions is derived based on the unique structural connectivity of each individual (Fig. S1). When multiple WCOs are coupled, the resulting patterns of synchronization are highly dependent upon the topology of the coupling, ensuring that these BNMs are highly sensitive to individual variability in the underlying anatomical connectivity \cite{Muldoon2016,Bansal2018b,Bansal2018a}. As illustrated in Fig. \ref{fig:conceptual_schematic}b and Fig. S2, we then perform \textit{in silico} experiments by systematically applying computational regional (nodal) stimulation to the BNM and assessing the resulting patterns of synchronization that emerge based on subject-specific and region-specific variation in structural connectivity.

\subsection*{Classification of brain states using cognitive systems.}

Traditionally, one would measure system-wide synchronization by calculating a measure such as the Kuramoto order parameter within the entire network of oscillators \cite{Kuramoto1975}.  However, in order to study how the stimulation of different brain regions drives brain function, we instead focus on the relationship between patterns of synchronization between large-scale cognitive systems. Thus, based on previous research \cite{Muldoon2016,Power2011}, we assigned each of the 76 brain regions into one of nine cognitive systems (Fig. S3). Each cognitive system is defined by regions that coactivate in support of a generalized class of cognitive functions. This delineation includes several sensory-motor related systems, including auditory (Aud), visual (V), and motor and somatosensory (MS) systems, as well as the ventral temporal association system (VT) that encapsulates regions involved in knowledge representation. Several of the systems are involved in functional roles that are generic across cognitive performance, including the attention system (Att), the medial default mode system (mDM), and two systems associated with cognitive control, cingulo-opercular (CP) and frontoparietal (FP) systems. Finally, we include the subcortical system (SC) that consists of the regions responsible for autonomic and primal functions. %regions from the "old" brain.

\begin{figure}
\centering
\includegraphics [width=\linewidth, keepaspectratio]
{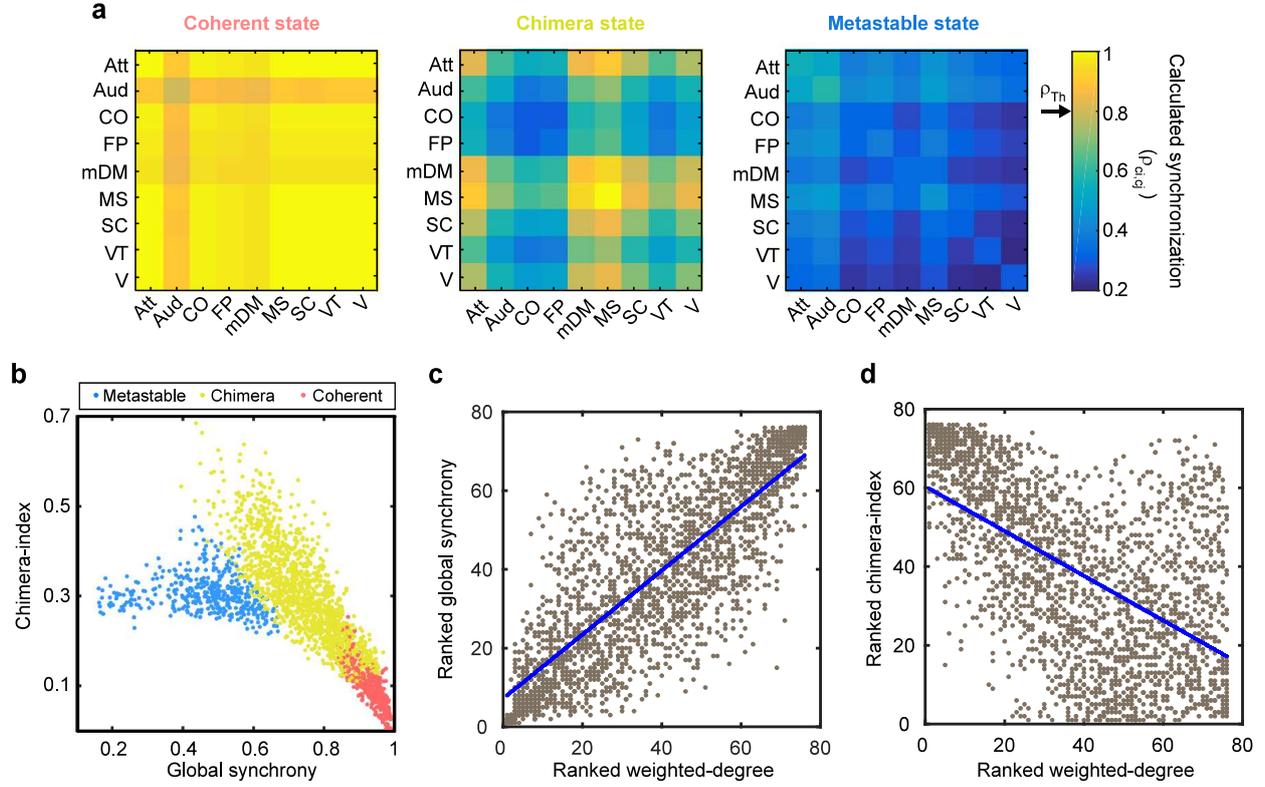}
\caption{Emergence of dynamical states within a cognitively-informed framework. (a) Cognitive system-level synchronization matrices whose entries denote the extent of synchronization between cognitive systems for coherent, chimera and metastable states. (b) The global synchronization and chimera-index of the system after stimulation of each brain region across all subjects. Within this traditional framework, our cognitively-defined states can be identified to have distinguishable characteristics. A coherent state shows high global synchronization value and low chimera-index (red). Both chimera (yellow) and metastable (blue) states show lower global synchronization. A chimera state can have either a higher global synchronization or a higher chimera-index than metastable states. (c)-(d) Origin of these states follows the connectivity of the stimulated brain region. (c) The global synchronization measure is positively correlated with the weighted degree of brain regions ($r=0.81$, $p=1.8\times 10^{-124}$), indicating that the network hubs are more likely to produce a coherent state. (d) The Chimera-index is weakly and negatively correlated with the weighted degree ($r=-0.57$, $p=7.4 \times10^{-195}$), indicating that stimulation of lower degree nodes is more likely to produce an ideal chimera state with half of the population synchronized and the other half de-synchronized.} 
\label{fig:states}
\end{figure}

In each \textit{in silico} experiment, we stimulate a brain region in each subject-specific BNM and then compute cognitive-system-level synchronization. We calculate a \textit{cognitive-system-based} Kuramoto order parameter $\rho_{c_i,c_j}$ that measures the amount of synchrony among all oscillators (regions) within two cognitive systems $c_i$ and $c_j$, and obtain a cognitive-system-based synchronization matrix as shown in Fig. \ref{fig:states}a. In this synchronization matrix we define two cognitive systems, $c_i$ and $c_j$, to be synchronized if $\rho_{c_i,c_j}$ exceeds a threshold value ($\rho_{Th}$). In Fig. \ref{fig:states}a we chose $\rho_{Th}=0.8$ to define three dynamical states observed in this study: (i) a coherent state, where all cognitive systems are synchronized; (ii) a cognitive chimera state, where some cognitive systems form a synchronized cluster (yellow) while the other systems remain incoherent (blue); and (iii) a metastable state, where no stable synchrony between cognitive systems is observed. 

%Thus, as predicted from previous small-scale neuronal modeling approaches \cite{Omelchenko2013,Vuellings2014,Hizanidis2014,Bera2016b}, our results confirm that chimera states emerge in large-scale brain dynamics. 

Next, we compare our \textit{cognitive-system-based} analysis with two traditional measures of synchronization: (1) the classical Kuramoto order parameter \cite{Kuramoto1975} calculated across all 76 oscillators (regions) that captures the level of global synchrony in the network, and (2) the chimera-index \cite{Shanahan2010,Hizanidis2016} that describes the closeness of the state with an ideal chimera state (see Methods). In Fig. \ref{fig:states}b, we show how the three dynamical states (coherent, chimera, and metastable) observed after stimulation of different brain regions in different individuals are distributed in the global synchrony and chimera-index parameter space. We observe a clear separation of these states, and as expected, global synchrony decreases from the coherent to chimera to metastable state. Thus, a cognitively-informed, systems-based classification of dynamical states is comparable to the traditional measures of estimating synchrony within a network. It is also robust across a range of threshold values ($\rho_{Th}$) as shown in Figures S4 and S5.

We also examine how these two traditional metrics of synchronization relate to the connectivity properties of the node (region) itself. As seen in Fig. \ref{fig:states}c, the level of global synchrony is positively correlated with the degree of the region being stimulated ($r=0.81; p=1.8 \times 10^{-124}$). Stimulation of a network hub (highly connected brain region) is therefore more likely to produce a coherent state, while stimulation of a non-hub is more likely to result in either a chimera or metastable state. Interestingly, Fig. \ref{fig:states}d reveals that the chimera-index shows a weaker and negative correlation with degree ($r = -0.57$; $p=7.4 \times 10^{-195}$). This relationship indicates the ability of moderately connected brain regions to produce a variety of spatially distinct synchronization patterns as a result of stimulation. These results not only demonstrate that chimera states emerge among large-scale cognitive systems but also reveal that the variable structural connectivity of regions within cognitive systems can drive the whole brain into diverse synchronization patterns.

\subsection*{Variable brain states emerge from stimulation of different brain regions.}

\begin{figure}
\centering
\includegraphics [width=\linewidth, keepaspectratio]
{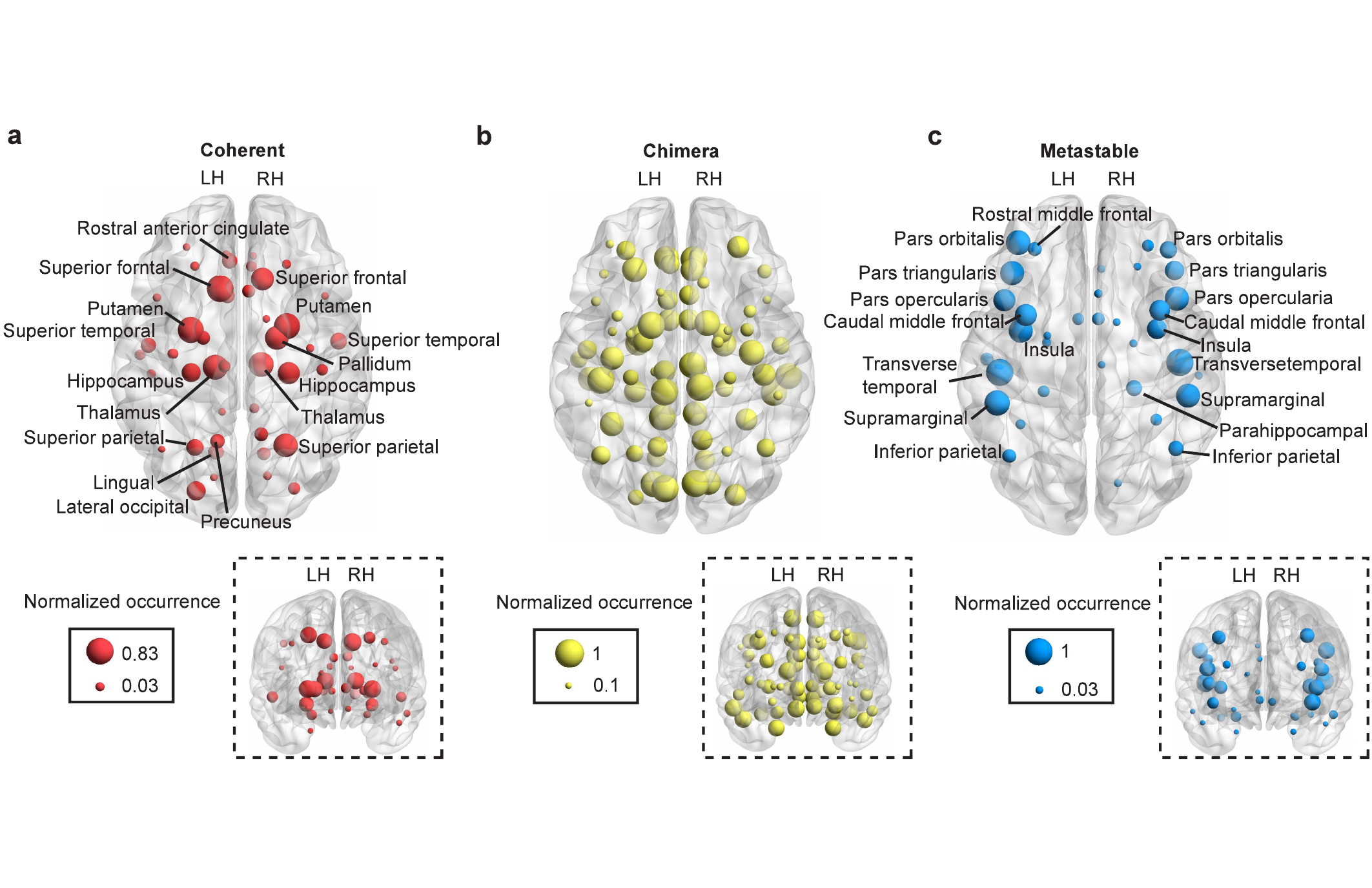}
\caption{Distributed origin of dynamical states. Separate depictions of brain network regions (nodes) that produce each of the three dynamical states: (a) coherent, (b) chimera, and (c) metastable. Both axial and coronal views of the brain are presented to visualize the anatomical location of the node in left (LH) and right (RH) hemispheres. Node size represents the normalized occurrence of a given state upon stimulation of the node across all the subjects. Variability in node sizes imply that different states can emerge upon the stimulation of a single node across individuals. A coherent state is significantly more likely to be originated from nodes within the center of the brain. (b) Chimera states are likely to be originated by the stimulation of nodes that are relatively equally distributed within the brain; however, these nodes may vary in the spatial patterns of synchronization that they produce. (c) A metastable state is more likely to originate from the stimulation of peripheral brain regions.}
\label{fig:states_distribution}
\end{figure}

Given that the type of brain state that emerges as result of stimulation is related to the local connectivity of the region, we next asked if there was also a spatial relationship between the location of the stimulated region and the type of dynamical state produced. In Fig. \ref{fig:states_distribution}, brain regions are depicted as an orb, and their sizes denote the normalized occurrence of a given state, that they produce upon stimulation, calculated across subjects. Regions that produce coherent states (network hubs) are distributed more closely to the center of the brain (\ref{fig:states_distribution}a), while regions that produce the opposite extreme, metastable states, are distributed along the periphery of the brain (\ref{fig:states_distribution}c). Interestingly, regions that produce chimera states are located between the two (\ref{fig:states_distribution}b). This asymmetry in the distribution of states highlights the differences in the structural organization of the brain. Regions in the center of the brain that produce a coherent state (e.g., subcortical regions such as the hippocampus, thalamus) can play a global cognitive role and facilitate communication between spatially separated brain regions.  Conversely, regions located in the periphery produce metastable states consistent with the notion that local and/or specialized computations take place in the cortex and could reflect the fact that the structure of the human brain has evolved to produce complex cognitive abilities \cite{Dehaene2015}.

\begin{figure}
\centering
\includegraphics [width=\linewidth, keepaspectratio]
{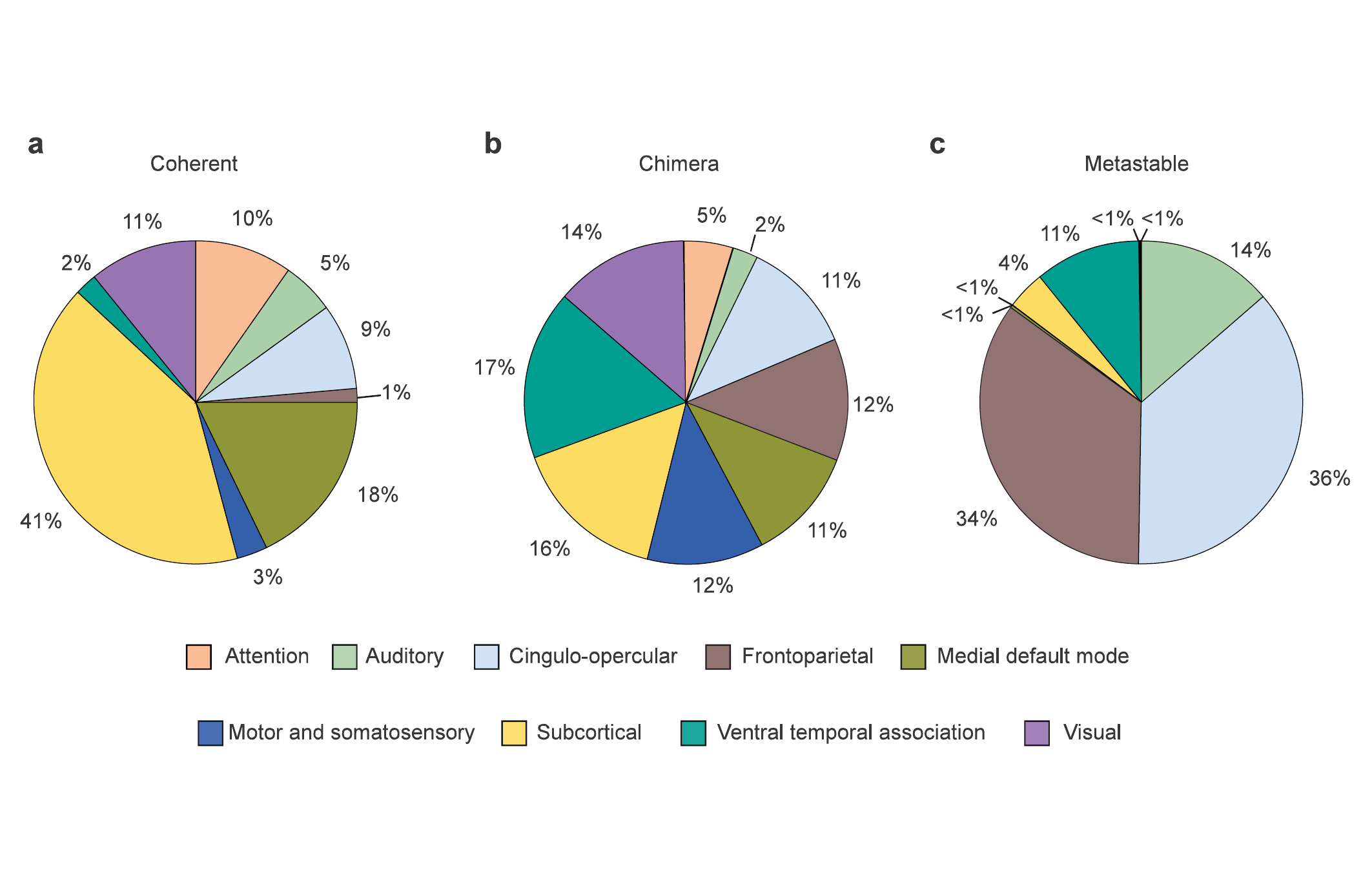}
\caption{Contribution of cognitive systems to dynamical states. (a) Coherent states are likely to result when nodes within medial default mode and subcortical systems are stimulated. (b) Chimera states emerge upon the stimulation of nodes that are  equally distributed across all the cognitive systems. (c) Metastable states frequently occur after stimulation of nodes within auditory, cingulo-opercular, frontoparietal, and ventral temporal association systems. This distribution indicates the dominance of a particular type of cognitive role within the nodes of different cognitive systems.}
\label{fig:systems_piechart}
\end{figure}

Due to the diversity in the distribution of dynamical states across spatially distributed regions, we investigate the relative contribution of the nine cognitive systems in producing each dynamical state after regional stimulation. As shown in Fig. \ref{fig:systems_piechart}a, coherent states are produced predominantly by regional stimulation within subcortical and medial default mode systems, reflecting that many of their constituent regions are network hubs. 

%With their extensive structural connections, hubs can drive the brain to full synchrony in \textit{in silico} experiments since the model does not include ongoing functional dynamics that can inhibit for more selective excitation among the structural connections.

There is also system specificity for metastable states which are preferentially produced by four systems (Fig. \ref{fig:systems_piechart}c). Two of those systems: cingulo-opercular and forntoparietal, are associated with cognitive control and proposed to be complementary systems that may often need to process task-relevant information concurrently. Their dominance in producing metastable states likely reflects the fact that these systems can work in seclusion without co-activating a large part of the brain (facilitating parallel processing) and that they are flexible and not constrained by their structural connectivity. Both the auditory and the ventral temporal association systems also contribute significantly to metastable states. These two systems are both predominantly located in the temporal lobe, an area of the brain associated with knowledge representation, so their functional roles may also frequently require working in seclusion in support of higher order perception.

Although coherent and metastable states are dominantly produced by specific cognitive systems, all nine systems give rise to chimera states (Fig. \ref{fig:systems_piechart}b). In fact, chimera states are more likely to occur than either coherent or metastable states (Figures \ref{fig:states_distribution}b and \ref{fig:states}b) and encompass a variety of different spatial patterns of coexisting coherent and incoherent behavior. This likely reflects the foundational role of partial synchrony in large-scale brain function. Cognitive tasks constantly require the intricate balance between segregated and integrated neural processing \cite{Shine2016}. The robust occurrence of chimera states following stimulation to each of the nine cognitive systems reflects the complexity and flexibility of the brain’s underlying architecture to support diverse processing requirements.

\subsection*{Structural variability influences observed dynamic states.}

\begin{figure}
\centering
\includegraphics [width=1\linewidth, keepaspectratio]
{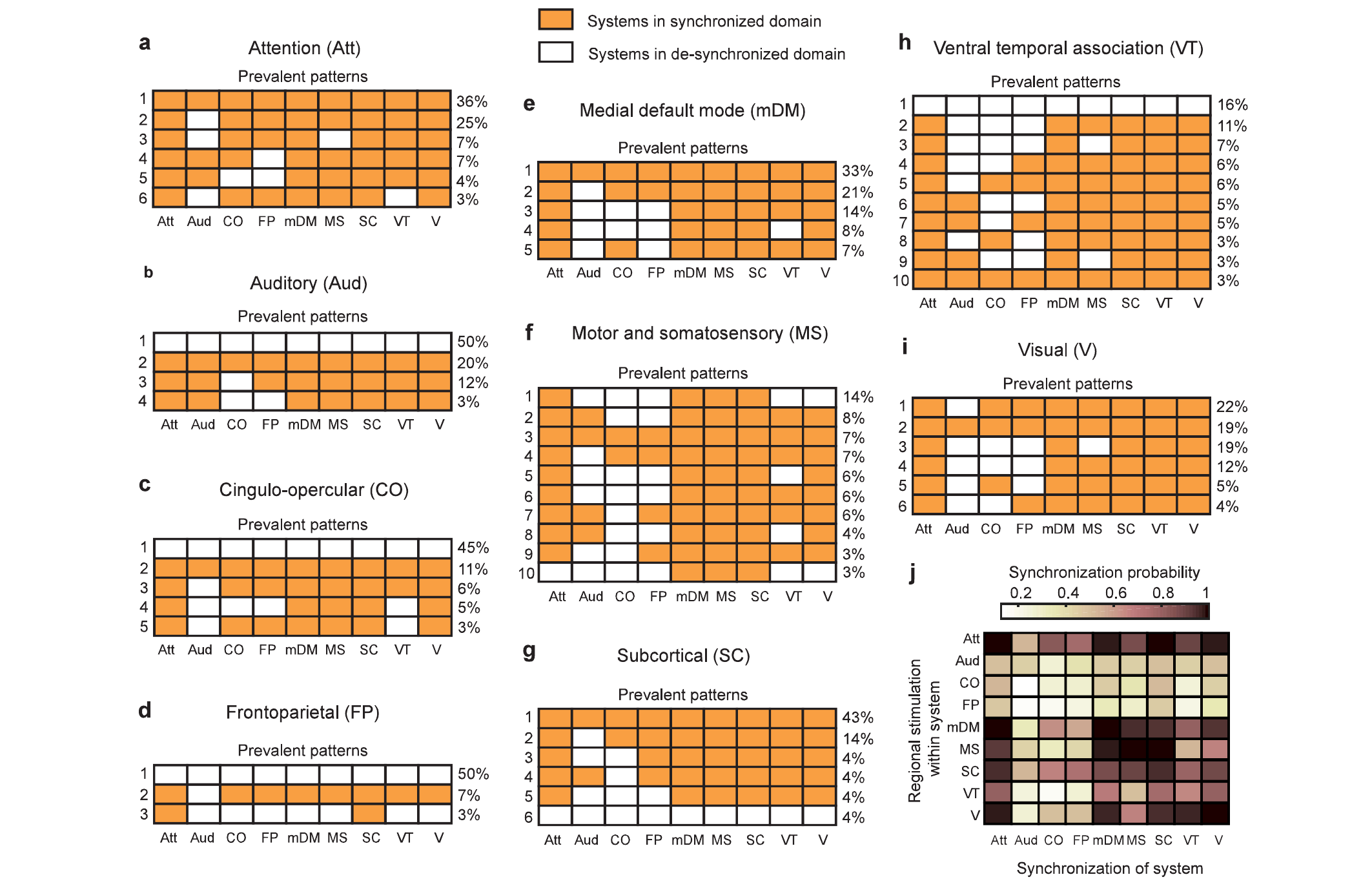}
\caption{Patterns of synchronization and cognitive chimera states. (a)-(i) Prevalent patterns (with an occurrence of at least 3\%) that emerge as the regions within different cognitive systems are stimulated across all subjects. Each panel represents stimulation of regions within a particular cognitive system. Each row represents one pattern of synchronization, and each column represents the state of a cognitive system. Cognitive systems that belong to the synchronized population are colored orange and cognitive systems that remain desynchronized are colored white. Thus, a fully orange or white row represents a coherent or metastable state respectively. Chimera states show different patterns of coloring depending on the cognitive systems that are recruited to the synchronized group. Different rows of patterns are stacked based on their relative occurrences (mentioned on the right side). To summarize the observed patterns, (j) shows the probability with which different cognitive systems can be synchronized when the regions within a given system are stimulated across subjects (shown along the vertical axis).}
\label{fig:patterns}
\end{figure}

We next characterize the spatial patterns that comprise the chimera states to understand which cognitive systems are synchronized and de-synchronized following stimulation to each region. The results are organized by stimulation of brain regions within a cognitive system in Fig. \ref{fig:patterns}a-i, where each row is a possible pattern of synchronization within the nine cognitive systems. The rows are organized by the frequency with which the pattern was observed (listed to the right of the row) after stimulation to all regions within that system and across all individuals in the study. For each pattern, systems that are part of the synchronized population after stimulation are shown in orange, and systems that are part of the de-synchronized population are shown in white. Consequently, coherent states are demarcated by a fully orange row, metastable states by a fully white row, and chimera states by a mixed pattern of coloring. For each cognitive system (Fig. \ref{fig:patterns}a-i), we present the prevalent patterns observed and these results illustrate what systems are likely to synchronize after stimulation.

Aligned with their complementary roles in cognitive control, the cingulo-opercular and frontoparietal systems continue to show similarity in their patterns of synchronization. The dominant pattern after stimulation to regions in either system is a metastable state, occurring 45\% for cingulo-opercular system (Fig. \ref{fig:patterns}c) and 50\% for the frontoparietal system (Fig. \ref{fig:patterns}d). Similarly, the auditory system also produces a metastable state 50\% of the time (Fig. \ref{fig:patterns}b). For all three of these systems, the second most frequent state is the opposite extreme, a coherent state (20\% for auditory and 11\% for cingulo-opercular) or nearly coherent state (7\% for frontoparietal). Thus, these three systems show diversity in the types of dynamical states that they are capable of producing: sometimes stimulation of the system produces a metastable (segregated) state, while other times, stimulation of the system drives the brain to a coherent (integrated) state.

The ventral temporal association system also produces a metastable state as its most prevalent pattern (Fig. \ref{fig:patterns}h), but unlike in the three previously discussed systems, this state is produced less frequently (16\%), and the system also produces a much larger variety of prevalent patterns of synchronization (10 unique patterns). The only other system to show this high level of diversity in its produced patterns is the motor and somatosensory system (10 patterns, Fig. \ref{fig:patterns}f). In both systems, multiple patterns of chimera states are observed. This likely reflects the ubiquitous need for neural processing related to both action coordination (motor and somatosensory) and higher order perception (ventral temporal association) to be integrated with the processing occurring in other systems within the brain \cite{Sheinberg2001,Spector2001}.

A coherent state occurs most frequently for the attention (36\%, Fig. \ref{fig:patterns}a), default mode (33\%, Fig. \ref{fig:patterns}e), and subcortical systems (43\%, Fig. \ref{fig:patterns}g). The visual system is also similar, though the coherent state is the second most prevalent (19\%) with a nearly coherent state (all but auditory) as its dominant pattern (22\%). Overall, these four systems are less dynamically diverse since their stimulation largely results in chimera states with high synchrony. All of these systems serve fundamental functional roles to rapidly respond to the external environment, and their dynamical patterns reflect this need to efficiently integrate this information with other cognitive systems.

For all systems, the most common state following regional stimulation is a chimera state with high synchrony, emphasizing the importance of partial synchrony for all of the diverse functional roles provided by large-scale cognitive systems. Visual inspection of the patterns suggests that the systems most likely to belong to the desynchronized population of a chimera state are the three systems that predominantly produce a metastable state (auditory, cingulo-opercular, and frontoparietal systems). This effect is quantified in the synchronization probability plot (Fig. \ref{fig:patterns}j). Systems with a high probability of synchronization have dark colors, while systems that are unlikely to be synchronized as a result of stimulation to a specific cognitive system are shown in light colors. While the auditory, cingulo-opercular, and frontoparietal systems systems are all unlikely to synchronize with other systems, the attention system, in contrast, is highly likely to be part of the synchronized population following stimulation to any system. We also observe that the stimulation of a region within a particular system does not necessarily induce synchronization within that system, which is particularly the case with the auditory, cingulo-opercular, and frontoparietal systems.

Collectively, these results reveal the power of this approach to characterize large-scale system effects after regional stimulation. In Fig. S6, we further describe how different nodes in a system contribute to observed patterns by spatially mapping the probability of a given pattern onto the brain. Taken together, these results highlight that stimulation within each system gives rise to multiple patterns, similar patterns can emerge from spatially different regions, and within a system, there can be a special distribution of states across brain regions. This likely arises from individual differences in the structural connectivity between the participants in the study or differences in the structural connectivity of the regions themselves within each system (or combination of the two). Consequently, we introduce a new metric to assess the contribution of subject-specific and region-specific variability on the observed patterns.

\subsection*{Dissociation of subject-specific and region-specific variability}

In our final analysis, we compute a measure called \textit{tenacity}, which we defined to quantify the level of similarity between a set of observed patterns (see Methods), to assess how structural variability, either between subjects or between regions, influenced the patterns observed in Fig. \ref{fig:patterns}. To differentiate these two potential sources of variability, we separately compute a subject tenacity and a region tenacity score (see Methods). When tenacity is calculated across subjects (subject tenacity), it measures the similarity of different patterns produced across individuals by stimulating the same brain region. A cognitive system's subject tenacity is then the average subject tenacity across regions within the system. When tenacity is calculated across brain regions (region tenacity), it measures the similarity of different patterns produced by stimulation of different brain regions within a given cognitive system in a single individual. This value is then averaged over all individuals. Consequently, a high value of tenacity indicates a high similarity between synchronization patterns produced by stimulation across individuals (subject tenacity) or brain regions (region tenacity).

\begin{figure}
\centering
\includegraphics [width=0.4\linewidth, keepaspectratio]
{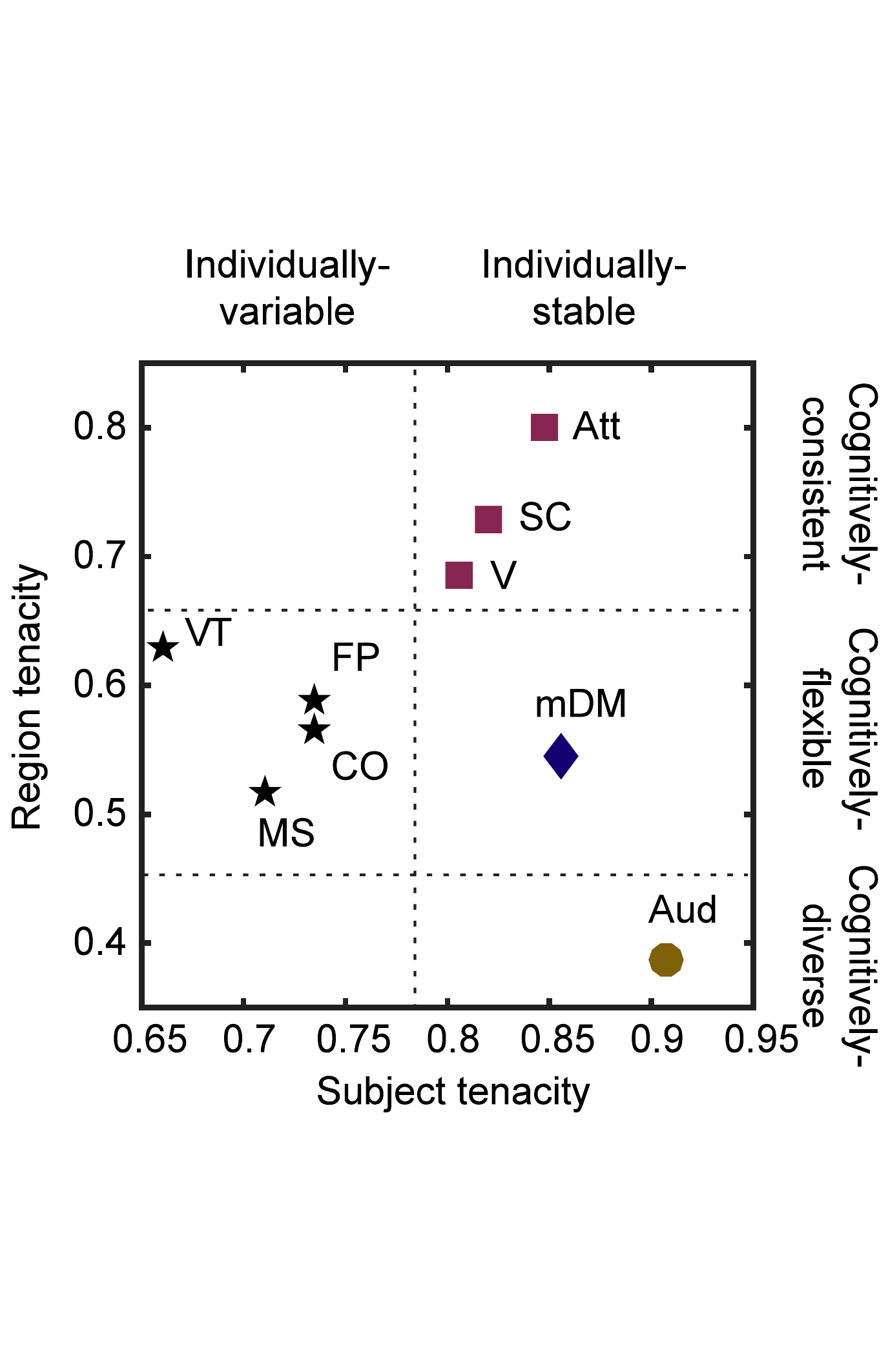}
\caption{Classification of cognitive systems based on pattern tenacity. To estimate the consistency of emergent patterns of synchronization within cognitive systems, we constructed a measure called tenacity that estimates the similarity between a set of patterns. Within a cognitive system, we calculate tenacity across two dimensions: patterns that are produced after stimulating each region across subjects (subject tenacity), and patterns that are produced after stimulating different regions of the system within each subject (region tenacity). In the parameter space constructed along these dimensions, we can cluster cognitive systems into $4$ groups that suggest a 2x3 partitioning of the tenacity space. This partitioning allows us to dissociate subject-specific and region-specific variability in observed patterns.}
\label{fig:tenacity}
\end{figure}

In Fig. \ref{fig:tenacity}, we plot each cognitive system based on its score for subject and region tenacity.  Additionally, we group cognitive systems by applying a clustering algorithm (see Methods), and the color and shape of a region's icon reflects its group assignment.  We identify four distinct groups of systems that are characteristically different from each other in terms of their location in the tenacity space and also in their cognitive roles.  To better delineate the four groups, we partition the tenacity space based on the level of subject and node tenacity.  We define two levels of subject tenacity: variable and stable, while the node tenacity is partitioned into three levels: diverse, flexible, and consistent.  

Across the subject tenacity dimension, we observe four individually-variable systems that demonstrate the largest variability in patterns. These systems include the frontoparietal and cingulo-opercular, the cognitive control systems that have previously been shown to have large individual variability \cite{Gratton2018}, and ventral temporal association and motor and somatosensory systems that show learning-dependent changes \cite{Muraskin2017,Sheinberg2001,Scholz2009}. The remaining five systems are classified as individually-stable with high subject tenacity scores. The default mode, subcortical, and attention systems have previously been found preserved across individuals as well as across species \cite{Margulies2006}, whereas the auditory and visual systems support fundamental perceptual processing \cite{Kaskan2005,Stevens2001}.

Across the region tenacity dimension, we observe three levels of tenacity. Congitively-consistent systems include the attention, subcortical, and visual systems, and stimulation to regions within these systems give rise to similar patterns of synchrony and de-synchrony. The auditory emerges as the sole cognitively-diverse system. This reflects the starkly different patterns that arise after stimulation of the regions within this system.  For example, stimulation to the superior temporal region results in high synchrony (both coherent and chimera) states, while stimulation to the transverse temporal region leads to a metastable state (see Fig. S6). Thus, the different local connectivity patterns of regions within this system produce immense diversity in the resulting synchrony patterns upon their stimulation. The remaining five cognitive systems are classified as cognitively-flexible, indicating that stimulation of regions within these systems produced variable patterns of synchrony.

Overall, the tenacity scores both confirm and extend our knowledge of brain structure-function relationships. The variability in subject tenacity among the systems reflects known differences in system stability between individuals, and confirms that variability in chimera patterns captures these coarse differences among the systems. On the other hand, the spread of region tenacity scores captures the diversity in the functional roles that the regions within a system serve across diverse cognitive tasks. Cognitively-consistent systems can largely be involved in the core sensory processing and associative learning, whereas the variability of patterns within cognitively-flexible systems may enable them to serve diverse cognitive roles, relying on stimulation of each constituent as a means to synchronize and integrate with different cognitive systems to support particular cognitive demand or task-relevant processing.

\section*{Discussion}

Using a novel, chimera-based framework, we explored the dynamical states that emerge across large-scale cognitive systems following the spread of a targeted regional stimulation. We identified three distinct dynamical states -- coherent, chimera, and metastable -- that arise as a function of the structural connectivity of the stimulated regions. A core result across all analyses is the variety in frequency and distribution of the observed dynamical states. Chimera states are the most pervasive state to emerge following regional stimulation. This likely reflects the foundational role that partial synchrony serves in large-scale brain function to enable the intricate balance between segregated and integrated neural processing. Furthermore, the diversity in these patterns captured both subject-specific and region-specific variability in structural connectivity. Based on its sensitivity to these different sources of variability, our novel chimera-based framework shows immense promise to better understand individual differences, relating patterns to performance, as well as system constraints that underlie how to drive the brain to different task-relevant states.

\subsection*{Prevalence of chimera states across cognitive systems enables segregation and integration in brain dynamics}
The brain is a complex dynamical system that must integrate information across spatially-distributed, segregated regions that serve specialized functional roles \cite{Alivisatos2012}. Neuroscience research therefore attempts to understand how the brain creates selective synchrony across subsets of task-relevant regions to enable rapid and adaptive cognitive processing \cite{Menon2010,Gollo2017}. Recently, network neuroscience approaches have identified sets of brain regions that form cognitive systems during rest and task states \cite{Smith2009}. By studying the interactions of these cognitive systems, functional analyses have identified the importance of (a) integrated states where the connections are stronger between cognitive systems\cite{Shine2016} and (b) segregated states where connections are weaker between cognitive systems and are likely to be stronger within. The relative level of functional integration vs segregation of cognitive systems has important consequences for cognitive performance. Highly segregated systems enable efficient computations in local, functionally-specialized brain regions, while strongly integrated systems provide rapid consolidation of information across systems necessary for coordinated, cohesive performance of complex tasks \cite{Shine2016,Vatansever2017,Bullmore2009}. These critical brain states are captured in the chimera framework as metastable (segregated) states and coherent (integrated) states, and perhaps the most critically for brain function, the chimera state that describes partial synchrony across subsets of cognitive systems.

All nine systems give rise to a chimera state following stimulation, suggesting that all cognitive systems can drive the brain to partial synchrony in support of their functional roles in cognition. Our results augment a burgeoning literature on what brain dynamics support rapid shifts between more segregated or integrated brain states. Previous work has found that functionally segregated states tend to involve shorter, local connections \cite{Liegeois2016}, while integration largely relies on the global influence of subcortical regions and cortical hubs that have many diverse connections to other brain regions \cite{Shine2018}. Collectively, our results demonstrate that our novel chimera-framework can investigate critical cognitive states where a balance between integration and segregation is required for adaptive cognition.

\subsection*{Chimera framework reveals subject-specific and region-specific variability in brain connectivity through the analyses of emergent dynamical states}

We found that a coherent state is likely to be produced by the stimulation of regions in the medial default mode and subcortical systems. This reflects the propensity of these systems to contain regional hubs, and the prevalent emergence of a coherent state reflects their functional roles to bridge spatially disperse regions and facilitate global brain communication. Interestingly, these systems also have high subject tenacity, indicating robust occurrence across the 30 individuals. This could reflect that the subcortical and medial default mode systems provide a fundamental, constant pillar of brain organization, which when disrupted, could lead to impairments in global brain function. Previous research has shown that network hubs are often found to be impacted in neurological disorders such as schizophrenia \cite{Klauser2017} and Alzheimer's disease \cite{Buckner2009}. These disorders are associated with network-wide deficits in brain function \cite{Menon2011}, which is consistent with our finding that cognitive systems that produce coherent states also contain network hubs. 

%In addition to the subcortical and medial default mode systems, our results revealed that the two sensory systems (auditory and visual) and the attention system have high subject tenacity. These systems provide the foundational processing for perceiving and orienting to external stimuli, which is critical for survival, and therefore, our results confirm the expectation that these systems would be highly stable across individuals \cite{Spector2001,LANDAU2009}. We also observed that the attention system is the most likely to be synchronized after stimulation, and this likely reflects the critical role of selective orienting and mental focus for cognition generally \cite{Arns2013}. 

Conversely, metastable states are preferentially produced by four systems with more sparse structural connectivity: two systems associated with cognitive control (cingulo-opercular and frontoparietal systems) and two systems associated with intricate sensory, object, and language representations (auditory and ventral temporal association systems). These systems all have functional roles that frequently require working in seclusion from other specialized processing in the brain. Interestingly, the three systems that are the most unlikely be synchronized upon stimulation are also the ones that are most likely to produce a metastable state: the auditory, cingulo-opercular, and frontoparietal systems. The cingulo-opercular and forntoparietal systems are associated with cognitive control, and they are proposed as complementary systems that are specialized for guiding successful task performance at different timescales: the cingulo-opercular system for maintaining task-related goals across trials, and the frontoparietal systems for trial-by-trial control \cite{Cocchi2013}. These functional roles may often need to occur concurrently, and their production of metastable states could indicate that these systems can work in seclusion without co-activating a large part of the brain, an attribute that facilitates parallel processing. 

%Furthermore, the prevalence of metastable states after stimulation denote that these systems are flexible and not constrained by their structural connectivity. While intriguing as an interpretation, future research is needed to examine the origin of these observations. What we identify as the metastable state could include synchronization at a smaller population level that our current framework is not sensitive to since we examine patterns of synchronization within cognitive systems (and not at the regional level). A more detailed modeling approach at a finer resolution could address this issue and is an important direction for future research.

While our chimera framework revealed stable features of brain architecture, it also captured cognitive systems where between subject variability leads to variety in frequency and type of synchronization patterns: cingulo-opercular, frontoparietal, ventral temporal association, and motor systems. These four systems with low subject tenacity are associated with higher cognitive functions where an individual’s experience and knowledge are likely captured by variability in their structural connectivity \cite{Powell2018}. Our results demonstrated that frontoparietal and cingulo-opercular systems exhibit the strongest individual variability, and this mirrors recent results that showed cognitive control systems have weaker within-subject variability and greater between-subject variability relative to sensory processing systems \cite{Gratton2018}. Our results also demonstrated that the ventral temporal association and motor/somatosensory systems show an especially high number of prevalent patterns with no single dominant pattern, and this may reflect their roles in learning and development-related changes \cite{Scholz2009,Kahn2017}.

\subsection*{Methodological Considerations and Future Directions}
Our model is only sensitive to functional relationships that are induced through structural connections, so the observed dynamical states and patterns are only constrained by the anatomical structure of the network. In reality, neuronal activity patterns that are observed in brain using different functional measurement techniques, such as fMRI, EEG, MEG, and PET, are a result of a complex neurophysiological activity that develops on top of the structural connectivity infrastructure. Thus, the actual patterns of brain activity that are observed across functional modalities may come from the simultaneous activation of different brain regions via multiple input sources and therefore might differ from the patterns observed in our \textit{in silico} experiments. Here, the emergence of a coherent pattern would imply that a node can, in principle, communicate with all of the spatially distributed regions within the brain; however, the actual nodes that it communicates with may vary between different tasks according to the specific cognitive demands of the task. Likewise, the emergence of a metastable state in our model may not reflect total de-synchrony, but just synchronization at a smaller population level that our framework simulates, requiring future models to study dynamical states at a finer spatial resolution. Despite these limitations, our approach is sensitive to variability in region-specific and subject-specific brain connectivity, and it can be used to answer fundamental questions concerning the cognitive organization of the human brain necessary for quantifying meaningful individual differences in brain architecture, supporting individualized medicine, performance enhancement technologies, and personalized stimulation protocols for treatment and/or augmentation. 
%add our excitement about the extension of this work to apply chimera analysis on functional neuroimaging data... Can we do that In 1-2 sentences here, it sounds interesting?

\subsection*{Conclusion}
By employing the cognitive system framework \cite{Power2011}, our novel chimera-based approach keeps the computational complexity of the analysis to a minimum while confirming the existence of chimera states in the large-scale cognitive systems in the human brain, as predicted from low-level, small scale models \cite{Omelchenko2013,Vuellings2014,Hizanidis2014,Bera2016b}. The partial synchrony observed in a chimera state has a natural link to the well-documented role of functional segregation and integration of cognitive systems thought to support cognition \cite{Menon2010,Deco2015}, and the approach captures robust system differences for those that are largely stable across people as well as those that capture individual training and expertise. Thus, our approach provides a rich opportunity to study how dynamical states give rise to variability in cognitive performance, providing the first link between how chimera states may subserve human behavior.

%\begin{methods}

\section*{Methods}

\subsection*{Human diffusion spectrum imaging data acquisition and pre-processing.}
Diffusion MRI analysis was performed on the 30 individual participant scans previously reported elsewhere \cite{verstynen2014organization}. Twenty male and ten female subjects were recruited from Pittsburgh and the Army Research Laboratory in Aberdeen, Maryland. All subjects were neurologically healthy, with no history of either head trauma or neurological or psychiatric illness. Subject ages ranged from 21 to 45 years of age at the time of scanning (mean age of 31 years) and four were left-handed (2 male, 2 female). All participants signed an informed consent approved by Carnegie Mellon University and conforming with the Declaration of Helsinki and were financially remunerated for their participation.

Macroscopic white matter pathways were imaged using a Diffusion Spectrum Imaging (DSI) aquisition sequence on a Siemens Verio 3T MRI system located at the Scientific Imaging \& Brain Research Center (SIBR) at Carnegie Mellon University using a 32-channel head coil. A total of 257-direction were sampled using a  twice-refocused spin-echo sequence (51 slices, TR = 9.916s, TE = 157ms, 2.4 x 2.4 x 2.4mm voxels, 231 x 231mm FoV, and b-max = 5,000s/mm2). Diffusion data were reconstructed using q-space diffeomorphic reconstruction (QSDR;\cite{yeh2011ntu}) with a diffusion sampling length ratio of 1.25 and a 2mm output resolution. 

\subsection*{Construction of individual structural brain networks.}

Whole-brain structural connectivity matrices were constructed for each subject using a bootstrapping approach. To minimize the impact of bias in the tractography parameter scheme on streamline generation, whole-brain fiber tractography \cite{yeh2013deterministic} was performed 100 times for each participant, generating 250,000 streamlines per iteration. Across the 100 iterations, values were randomly sampled for QA-based fiber termination thresholds (0.01-0.10), turning angle thresholds (40 Ì-80 Ì), and smoothing (50\%-80\%), while constant values were used for step size (1mm) and min/max fiber length thresholds (10mm/400mm). On each iteration a binary connectivity matrix was generated, where an edge between two regions of interest was considered present if 5\% or more of streamlines generated were found to connect them. The probability of observing a connection was estimated by calculating the frequency of detecting an edge across all 100 iterations. The region of interest were determined using cortical components of the Desikan-Killiany atlas and subcortical components of the Harvard-Oxford Subcortical atlas. In the resulting weighted matrices, connection strengths were normalized by the sum of the regional brain volumes, and these normalized matrices were used as the structural representations of individual brains.

All analysis was performed using DSI Studio (http://dsi-studio.labsolver.org/) and Matlab (MathWorks, Inc.; Natick, MA, USA).

\subsection*{Data-driven network model of brain dynamics.}
In our data-driven network model, regional brain dynamics are given by Wilson-Cowan oscillators \cite{{Wilson1972},{Muldoon2016},{Bansal2018b}}. In this biologically motivated neural mass model, the fraction of excitatory and inhibitory neurons active at time $t$ in the $i^{th}$ brain region are denoted by $E_i(t)$ and $I_i(t)$ respectively, and their temporal dynamics are given by:  
\begin{equation}
\tau\frac{dE_i}{dt}=-E_i(t)+(S_{E_m} - E_i(t))S_E\Big(c_1E_i(t)-c_2I_i(t) \\+ c_5\sum\limits_{j}A_{ij}E_j(t-\tau_d^{ij})+P_i(t)\Big)+\sigma w_i(t),
\end{equation}
\begin{equation}
\tau\frac{dI_i}{dt}=-I_i(t)+(S_{I_m} - I_i(t))S_I\Big(c_3E_i(t)-c_4I_i(t) \\+ c_6\sum\limits_{j}A_{ij}I_j(t-\tau_d^{ij})\Big)+\sigma v_i(t),
\end{equation}
where
\begin{equation}
S_{E,I}(x) = \frac{1}{1+e^{(-a_{E,I}(x-\theta_{E,I})}} - \frac{1}{1+e^{a_{E,I}\theta_{E,I}}}.
\end{equation}

\noindent $A_{ij}$ is an element of the subject-specific coupling matrix $A$ whose value is the connection strength between brain regions $i$ and $j$ as determined from diffusion spectrum imaging as described above. The global strength of coupling between brain regions is tuned by excitatory and inhibitory coupling parameters $c_5$ and $c_6$ respectively. In this case $c_6=c_5/4$. $P_i(t)$ represents the external stimulation to excitatory state activity and was used to perform computational stimulation experiments.  The parameter $\tau_d^{ij}$ represents the communication delay between regions $i$ and $j$. If the spatial distance between regions $i$ and $j$ is $d_{ij}$, $\tau_d^{ij}=d_{ij}/t_d$, where $t_d = 10m/s$ is the signal transmission velocity. We added noise as an input to the system through the parameters $w_i(t)$ and $v_i(t)$ which are derived from a normal distribution with $\sigma = 10^{-5}$. Other constants in the model are biologically derived: $c_1 = 16$, $c_2 = 12$, $c_3 = 15$, $c_4 = 3$, $a_E = 1.3$, $a_I = 2$, $\theta_E = 4$, $\theta_I = 3.7$, $\tau = 8$ as described in references \cite{{Wilson1972},{Muldoon2016},{Bansal2018b}}.  
To numerically simulate the dynamics of the system, we used a second order Runge Kutta method with step size $0.1$ with initial conditions $(E_i(0),I_i(0)=0.1,0.1)$.  

\subsection*{Targeted stimulation.}
The model was optimized for each individual to allow a regime of maximum dynamical sensitivity. This was done by choosing a global coupling parameter, $c_5$, such that the system was just below the critical transition point to the excited state (see Fig. S1).  Regional stimulation was achieved by applying a constant external input $P_i = 1.15$ to a single region and perturbing its dynamics (Fig. S2). As the dynamics evolve, the stimulation spreads within the brain through the network connectivity of the stimulated node. 

\subsection*{Cognitive systems.}
We assigned each brain region to one of $9$ cognitive systems: attention, auditory, cingulo-opercular, frontoparietal, medial default mode, motor and somatosensory, subcortical, ventral temporal association and visual. This node assignment is described in Table S1 and is similar to the one used by Muldoon et al. \cite{Muldoon2016}. The distribution of brain regions within cognitive systems is shown in Fig. S3.

\subsection*{Calculation of synchronization within cognitively-informed framework.}
We used the standard order parameter $\rho$ to estimate the extent of synchronization after a targeted regional stimulation within the brain networks. This measure was proposed by Kuramoto for the estimation of coherence in a population of Kuramoto phase oscillators \cite{Kuramoto1975}. In this case, the instantaneous order parameter at a given time $t$ it is defined as 
\begin{equation}
\rho_N(t)e^{i\Phi(t)}=\frac{1}{N}\sum_{j=1}^{N}e^{i\phi_{j}(t)},
\end{equation}
 where $\phi_{j}$ is the phase of the $j^{th}$ oscillator at time $t$ and is given by 
\begin{equation}
\phi_j(t)=tan^{-1}\frac{I_j(t)}{E_j(t)}.
\end{equation}

\noindent Here, $N=76$ is the number of oscillators in the system. In order to estimate the global synchronization in the system, one needs to average the instantaneous order parameter for a sufficiently long period of time, 
\begin{equation}
\rho_N=<\rho_N(t)>_t.
\end{equation}
We used simulated activity over $1$ s to estimate the average order parameter.

Within our cognitively-informed framework, we measured the synchronization between all pairs of cognitive systems following a regional stimulation. This was done by calculating an order parameter for the combined oscillator population of a pair of cognitive systems. For cognitive systems  $c_i$ and $c_j$, this order parameter is given by
\begin{equation}
\rho_{c_i,c_j}=<\rho_{c_i,c_j}(t)>_t,
\end{equation}

\noindent where

\begin{equation}
\rho_{c_i,c_j}(t)e^{i\Theta(t)}=\frac{1}{N_{c_i}+N_{c_j}} \sum_{k\in (c_i\cup c_j)}e^{i\phi_{k}(t)}.
\end{equation}

\noindent Here, $N_{c_i}$ and $N_{c_j}$ represent the number of oscillators (brain regions or nodes) within cognitive systems $c_i$ and $c_j$ respectively. 

This analysis resulted in synchronization matrices, as shown in Fig. 2a, whose entries represent the extent of synchronization between cognitive systems. These matrices were used in order to identify the dynamical cognitive state that emerged as a result of regional stimulation.

%\subsection{Classification of cognitive states.}
%If all the cognitive systems are mutually synchronized ($ s_{ij} = 1 \forall (i,j)$), we identified the dynamical state to be a coherent state; if some of the cognitive systems form synchronized population, leaving others to be desynchronized, we identified the state to be a chimera state; and if none of the cognitive systems showed synchronization with any other cognitive system ($s_{ij} = 0 \forall (i,j)$), we identify the dynamical state to be a metastable state. We do not call a metastable state 'desynchronized' state for two reasons - firstly, we limit our calculation to the entire cognitive system, which can not distinguish a state where a small sub-population of the system is synchronized while a large population remains desynchronized resulting in an overall low value of the Kuramoto order parameter. Secondly, systems can produce instantaneous bursts of synchronous activity, which could be important for the functional dynamics of the brain, but will produce a low time averaged value of the Kuramoto order parameter in our framework. A metastable state therefore is the absence of large scale (both spatial and temporal) synchronization and we understand that the brain regions producing such a state are structurally constrained to produced localized activity and can have a cognitive role that is localized or specialized. 

\subsection*{Chimera-index.}
We calculated the chimera-index ($C$) as described in \cite{{Shanahan2010},{Hizanidis2016}} as a measure of the normalized average variation in order parameter within cognitive systems over time. For $ c_i \in [c_1,c_2,...,c_M$] ($M=9$ is the total number of systems),

\begin{equation}
C=\frac{<\sigma_{ch}(t)>_t}{C_{Max}},
\end{equation}
where
\begin{equation}
\sigma_{ch}(t)=\frac{1}{M-1}\sum_{i=1}^{M}(\rho_{c_i}(t)-<\rho_c(t)>_M)^2.
\end{equation}
\noindent In this case, $C_{Max}=5/36$  is a normalization factor and represents the maximum value of variability in the order parameter in an ideal chimera state where the network organizes such that the half of its population is completely synchronized and half is completely desynchronized \cite{Shanahan2010}. The instantaneous quantity $<\rho_c(t)>_M$ measures the synchronization of cognitive systems, averaged over all systems at a given time $t$.

\subsection*{Robust detection of the patterns of synchronization.}
In order to robustly identify emergent cognitive patterns, we first obtained a binarized synchronization matrix ($s$) such that, $s_{ij}=1$  if systems $i$ and $j$ are identified synchronized, and $s_{ij}=0$ otherwise. We defined two cognitive systems $i$ and $j$ to be synchronized if $\rho_{c_i,c_j}\geq\rho_{Th}$, where $\rho_{Th}$ represents a synchronization threshold. For the results discussed in the main text we used $\rho_{Th}=0.8$ \cite{Shanahan2010} (as indicated in Fig. 2a).

In principle, one can directly use such binarized synchronization matrices in order to classify the emergent states and patterns. However, we performed community detection on these binarized matrices. This method clusters the group of synchronized systems into a single community whereas desynchronized systems remain separate communities. In case of a coherent state, we observed only one community, and in case of a metastable state we observed nine separate communities, each representing a cognitive system. For chimera states, communities with different distributions of cognitive systems emerged. Thus, applying the community detection algorithm not only allowed us to robustly classify the emergent dynamical states, but also let us separate various spatially distributed patterns of chimera states. Community detection was performed using modularity maximization through the generalized Louvain algorithm \cite{Mucha2010}. For community detection, the value of the resolution parameter was varied between 0.8 to 0.95 and a consensus was run to determine the community structure \cite{Bassett2013}.

\subsection*{Pattern tenacity.}
A pattern in our analysis describes if the given cognitive system falls into the synchronized population or remains desynchronized (Fig. 5). In order to calculate similarity between patterns, we defined tenacity for a set of observed patterns as follows
\begin{equation}
T = \frac{1}{p(p-1)}\sum_{i,j=1}^{p}\big(\frac{1}{M}\sum_{c=1}^{M}\delta_{i,j}^{c}\big),
\end{equation}
where $p$ is the number of patterns in the set. $\delta_{i,j}^{c} =1$ if the cognitive system $c$ falls into the same state of either synchrony or desynchrony in patterns $i$ and $j$, and $=0$ otherwise.
We calculated the tenacity of cognitive systems in two dimensions: across subjects for a given brain region within the system (subject tenacity), and across regions of the system in a given subject (region tenacity). For subject tenacity of a given cognitive system, $p$ constitutes the patterns that are produced by a given node for all the subjects and equals the number of subjects i.e. $30$. For region tenacity of a given cognitive system, $p$ constitutes the patterns that all the nodes for a given cognitive system produce for a given subject, and the value of $p$ varies between systems. Thus for each cognitive system we obtained two distributions of tenacity, one for each subject and region tenacity.  

\subsection*{Clustering of cognitive systems using tenacity features.}
In the subject-node tenacity parameter space, we grouped cognitive systems into clusters using the k-means algorithm and silhouette analysis. We used $k =$ 3, 4, 5, and 6 and identified the stable clustering that maximizes similarity within clusters and dissimilarity across clusters. For $k=4$ we observed an optimized clustering. The corresponding silhouette plot is shown in Fig. S7.
 
\subsection*{Rendering of brain images.}
BrainNet Viewer was used to perform spatial mapping onto brain images \cite{Xia2013BrainNet}.

%\subsection*{Data availability.}
% Data will be made publicly available upon publication of the manuscript. 

%\end{methods}

%% Put the bibliography here, most people will use BiBTeX in
%% which case the environment below should be replaced with
%% the \bibliography{} command.

%\section*{References}

%\bibliography{references}

\section*{Acknowledgments}
This work is supported by the Army Research Laboratory through contract \# W911NF-10-2-0022 and W911NF-16-2-0158 from the U.S. Army research office. The authors acknowledge Gregory Lieberman for the pre-processing of the brain anatomical connectivity data, and Rajarshi Roy and Joseph D. Hart for a useful discussion on chimera states. The content is solely the responsibility of the authors and does not necessarily represent the official views of the funding agency.

\section*{Author contributions}
KB conceived the idea; KB, JMV, and SFM designed the research; TV and JMV collected the data; SFM developed the model; KB implemented the model and led the analysis; all authors contributed analysis ideas and wrote the paper.

\newpage
\beginsupplement

\begin{figure}
\centering
\includegraphics [width=0.7\linewidth, keepaspectratio]
{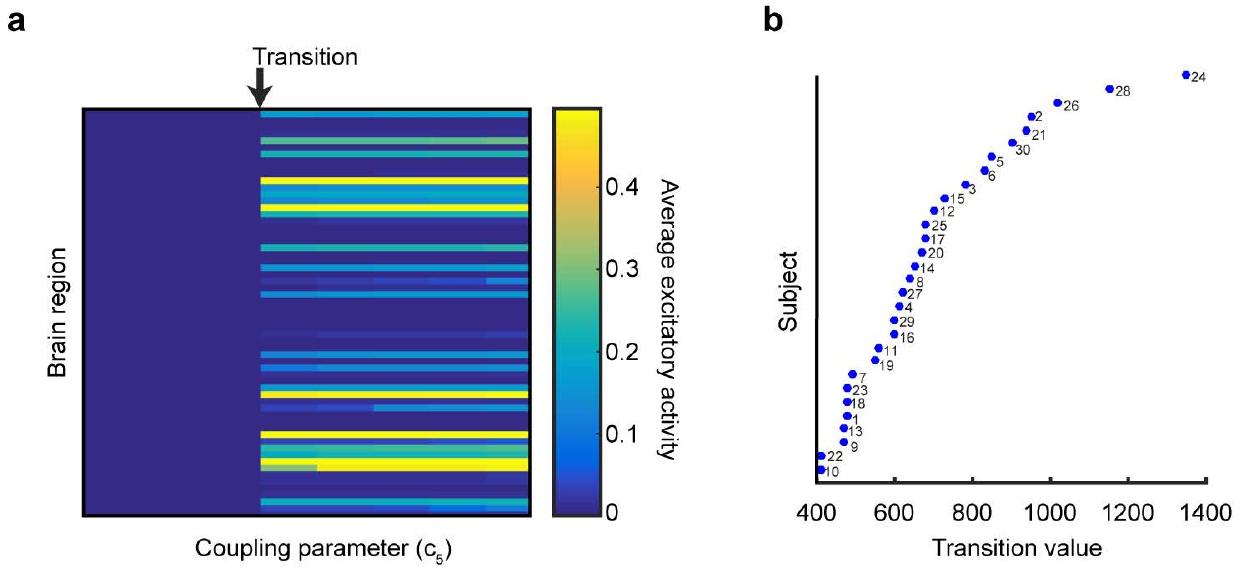}
\caption{Optimizing personalized brain network models. In our model, coupling between different brain regions can be tuned by the global coupling parameter,$ c_5$. We optimized this parameter for each individual to place the brain in a dynamical state that is maximally sensitive to a perturbation (stimulation). To find this optimal value for each individual, we varied $c_5$ between 100 and 1500 in steps of 10 and measured average excitatory activity ($E_i$) of each brain region ($i$) for a time period of 1 s. Each simulation was started with ($E_i$ , $I_i$) = (0.1, 0.1),  and before starting the measurement, we let the system evolve for 1 s in order to eliminate initial transients. (a) As we varied the value of c5, we observed a sudden transition in system’s behavior to an excited state. (b) The value of $ c_5$ at which the transition is observed typically varies between individuals. The transition value signifies the ease with which a given brain can be excited. In order to apply a computational stimulation within the personalized model, we fixed the value of $ c_5$ just below its transition value.}
\label{fig:s1}
\end{figure}

\newpage
\begin{figure}
\centering
\includegraphics [width=0.7\linewidth, keepaspectratio]
{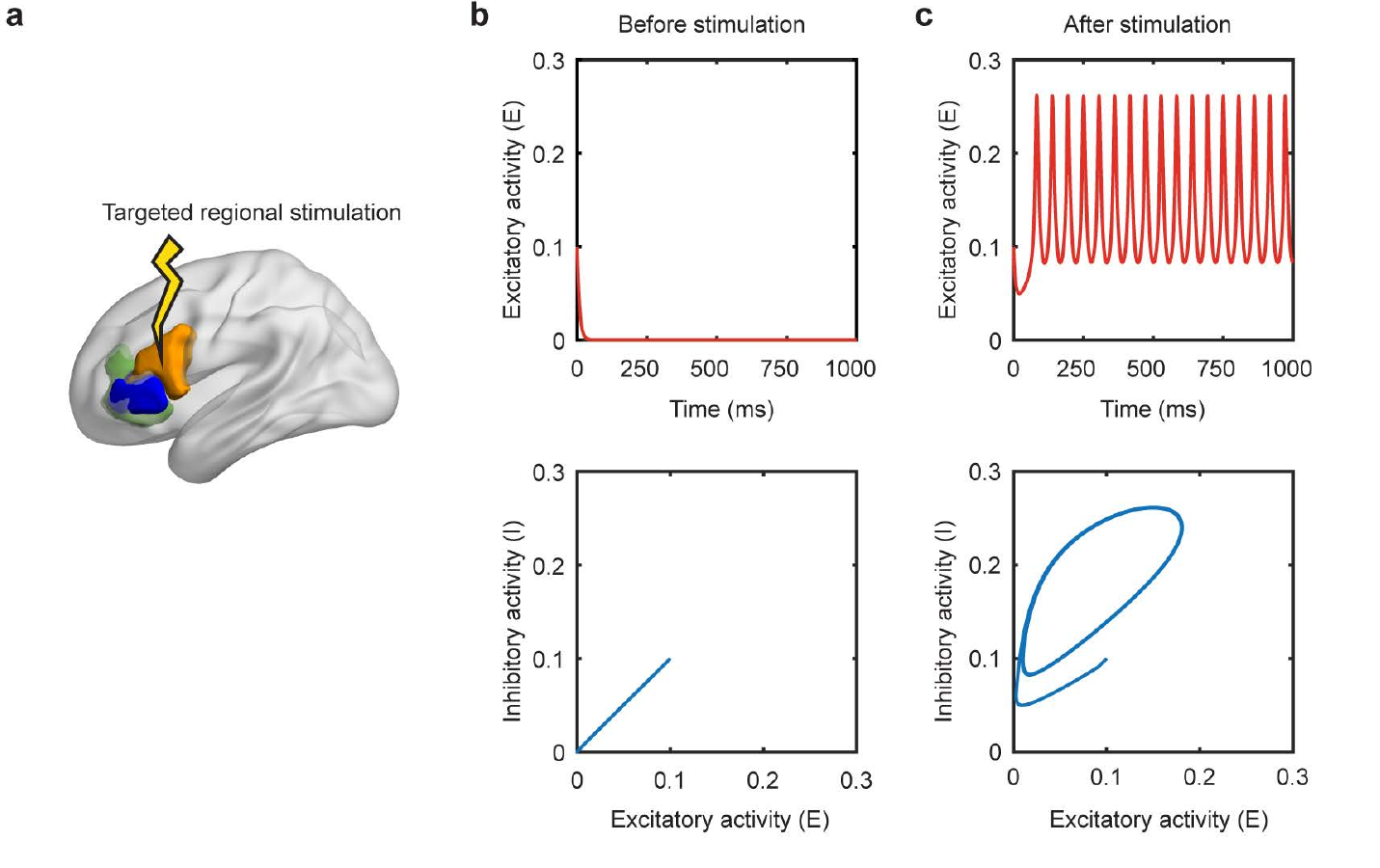}
\caption{Targeted regional stimulation. (a) To stimulate a particular brain region within the brain network model of a given individual, we applied a constant external input, $P_i = 1.15$. In our model, as we switch on the external input, the stimulated brain region changes its dynamics from a (b) stable fixed point to (c) a limit cycle (oscillatory motion). The effect of this stimulation on the other regions within the brain is measured through the resulting patterns of synchronization as discussed in the main text. }
\label{fig:s2}
\end{figure}

\newpage
\begin{figure}
\centering
\includegraphics [width=0.7\linewidth, keepaspectratio]
{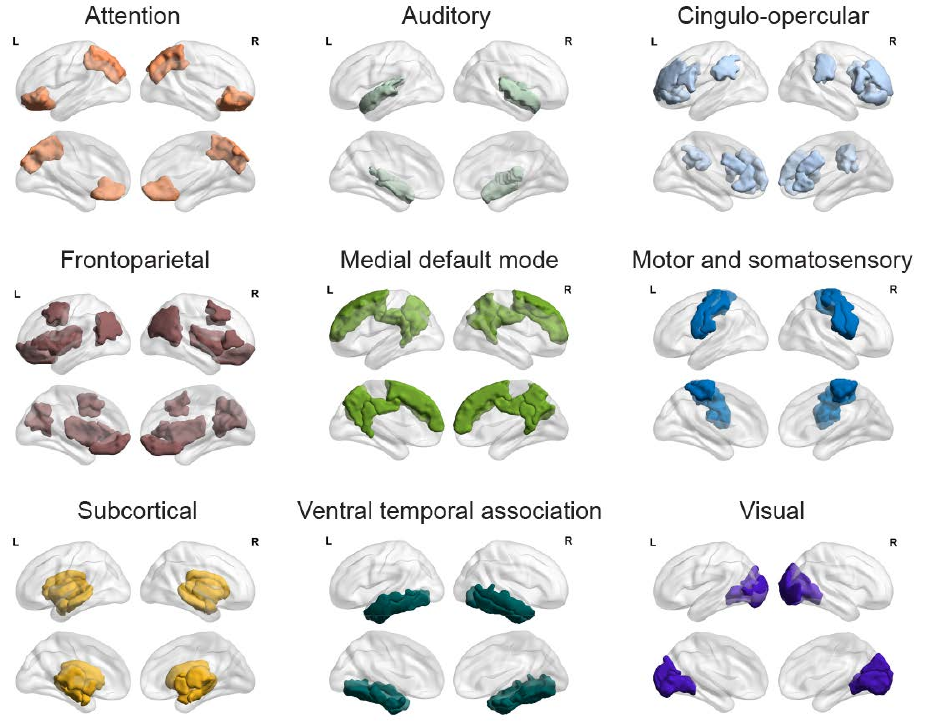}
\caption{Distribution of brain volume within cognitive systems. Spatial mappings showing the distribution of the nine cognitive systems within the brain.  The identification of regions is given in Table S1.  L and R denote the left and right hemisphere respectively.}
\label{fig:s3}
\end{figure}

\newpage
\begin{figure}
\centering
\includegraphics [width=0.7\linewidth, keepaspectratio]
{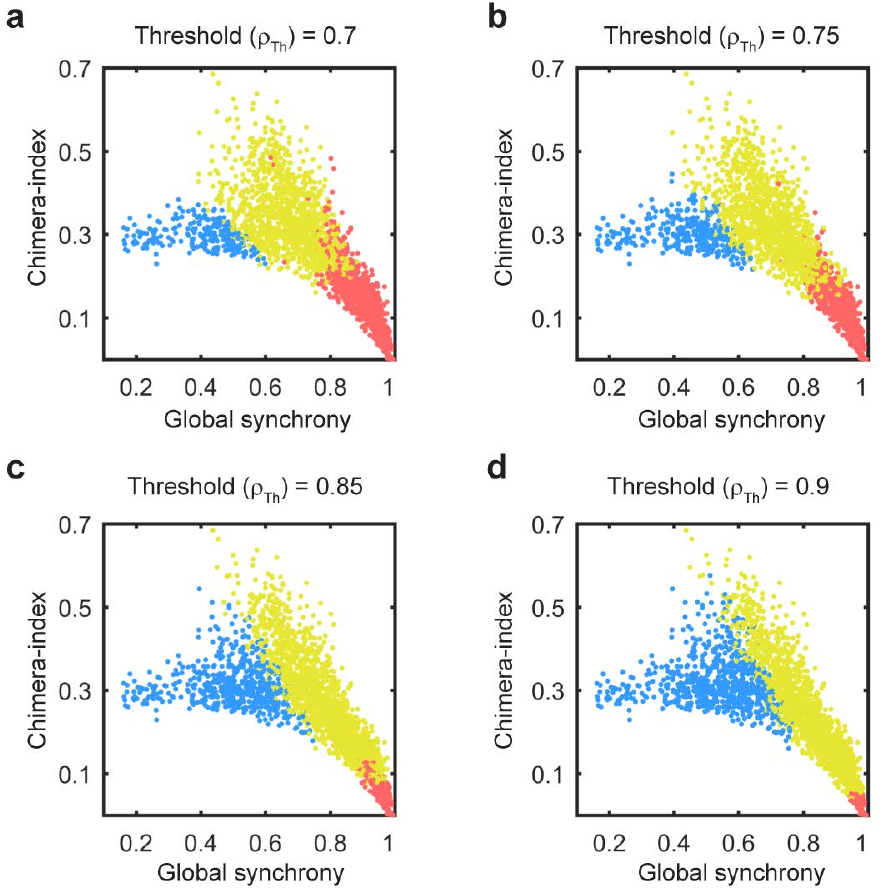}
\caption{ Effect of changing synchronization threshold on the distribution of states. (a)-(d) Separation of three different states in the parameter space of global synchrony and chimera-index for different synchronization threshold values ($\rho_{Th}$). Using different values of $\rho_{Th}$  in our analysis, we obtain results that are qualitatively similar to the results discussed in Fig. 2b. We observe three different states i.e. coherent, chimera and metastable states, for all four threshold values. These states can be clearly separated in the parameter space of global synchrony and chimera-index. However, a higher threshold value (e.g. 0.9), which signifies strictness in identifying a population to be synchronized, allows a lower number of states to be classified as a coherent state, as opposed to a lower threshold value (e.g. 0.7), which is more relaxed. Conversely, number of observed metastable states increases with increasing threshold values. }
\label{fig:s4}
\end{figure}

\newpage
\begin{figure}
\centering
\includegraphics [width=0.7\linewidth, keepaspectratio]
{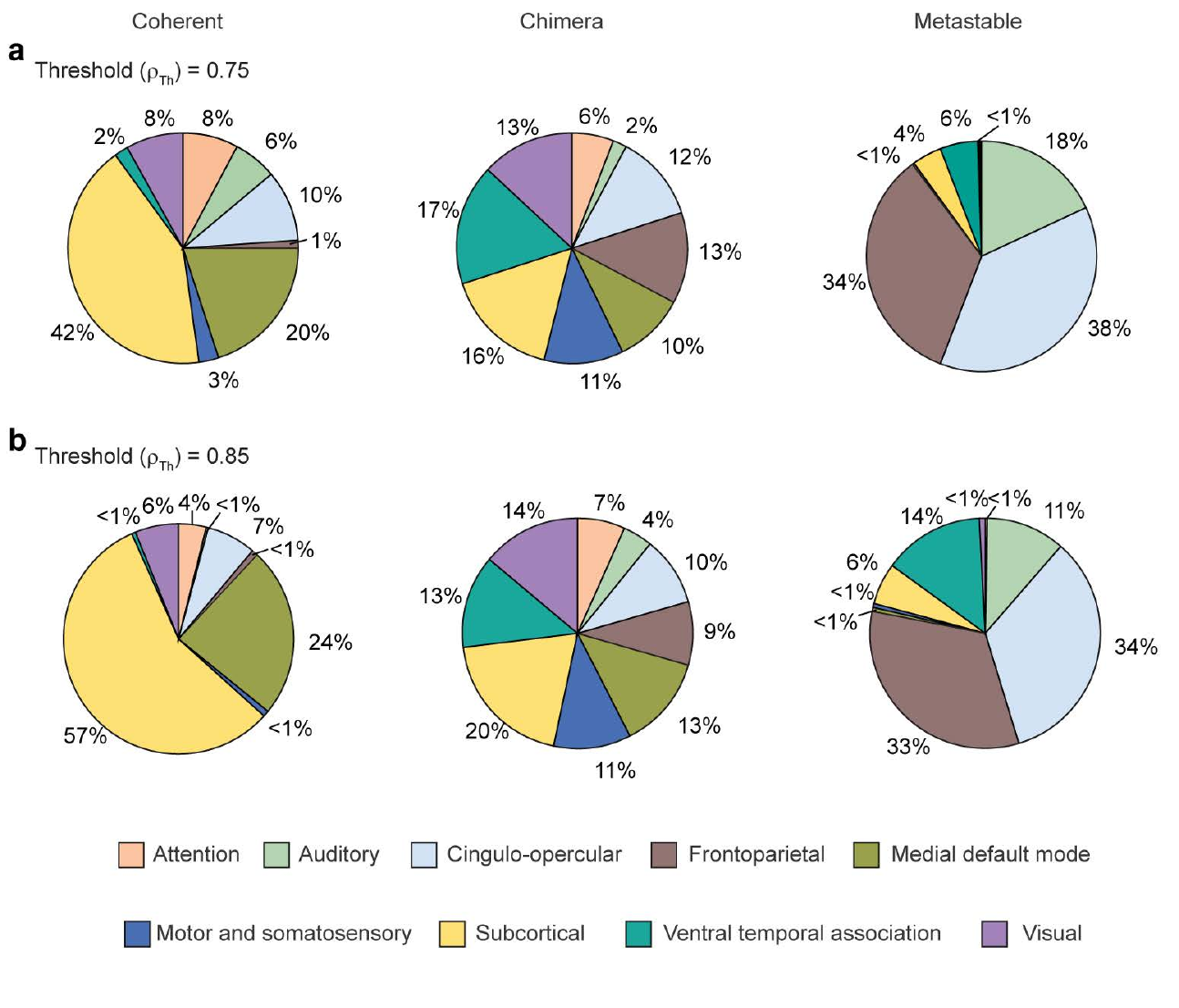}
\caption {Likelihood of the emergence of dynamical states across cognitive systems. Contribution of different cognitive systems towards producing a given cognitive states is depicted for (a) $\rho_{Th}=0.75$ and (b) $\rho_{Th} = 0.85$. These figures corroborate our findings presented in Fig. 4. We observe that a coherent state dominantly originates when the regions in subcortical and medial default mode systems are stimulated. A metastable state is likely to originate upon stimulation of the regions within cingulo-opercular, frontoparietal, auditory and ventral temporal association systems. Chimera states do not show the dominance of any particular cognitive system, regions from different cognitive systems are relatively equally likely to produce of chimera state upon stimulation.}
\label{fig:s5}
\end{figure}

\newpage
\begin{figure}
\centering
\includegraphics [width=0.7\linewidth, keepaspectratio]
{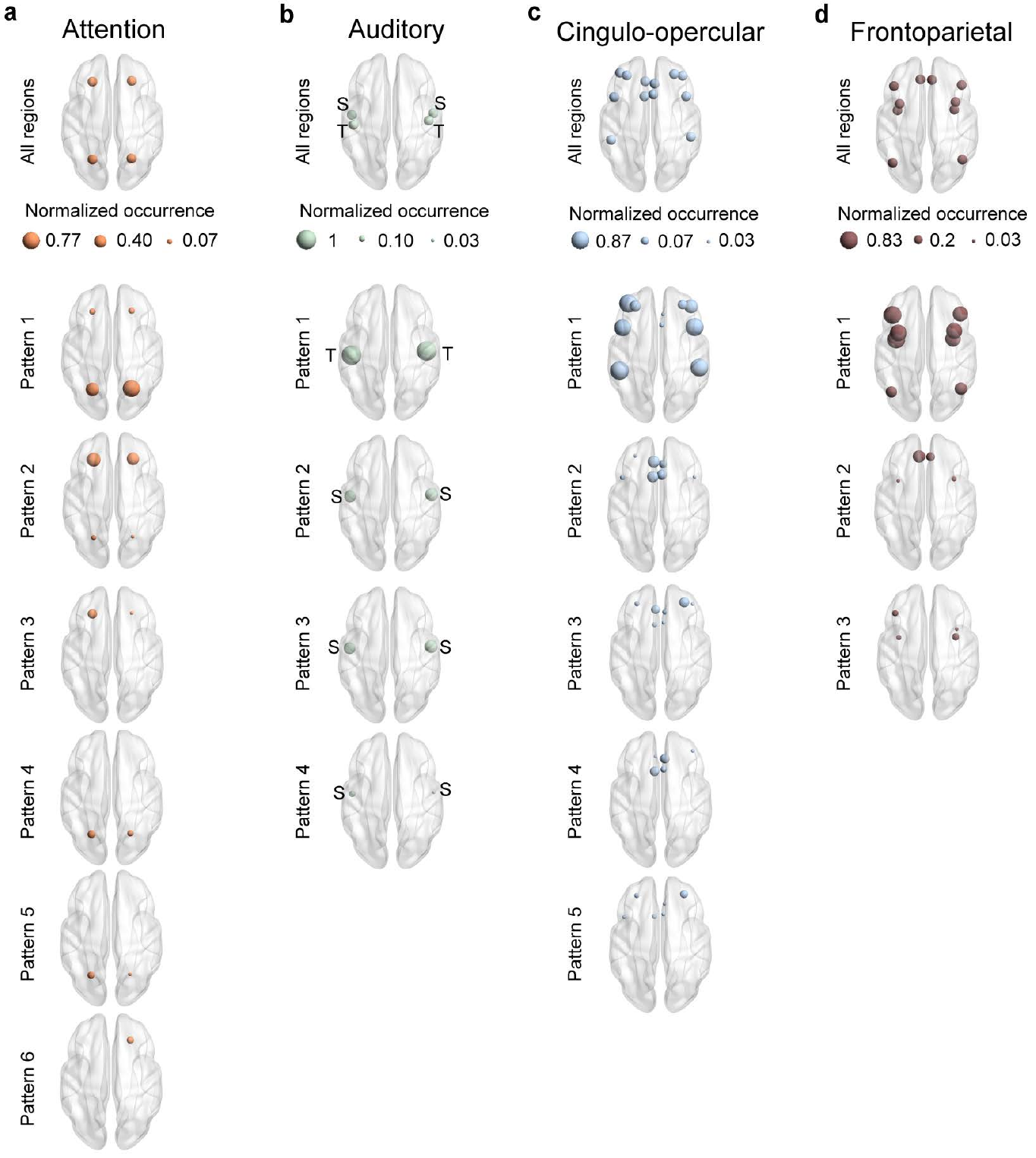}
\label{fig:s61}
\end{figure}

\newpage
\begin{figure}
\centering
\includegraphics [width=0.7\linewidth, keepaspectratio]
{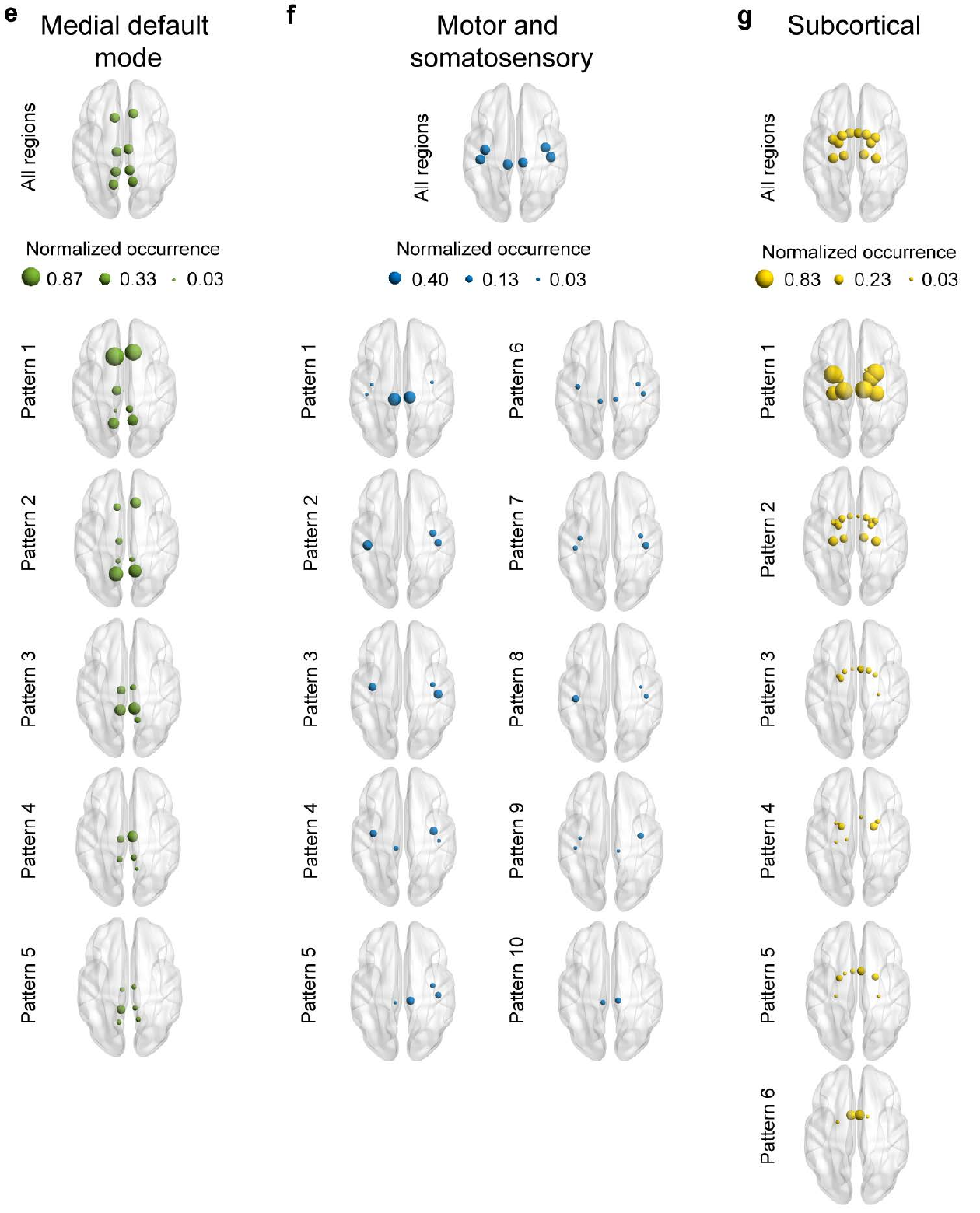}
\label{fig:s62}
\end{figure}

\newpage
\begin{figure}
\centering
\includegraphics [width=0.45\linewidth, keepaspectratio]
{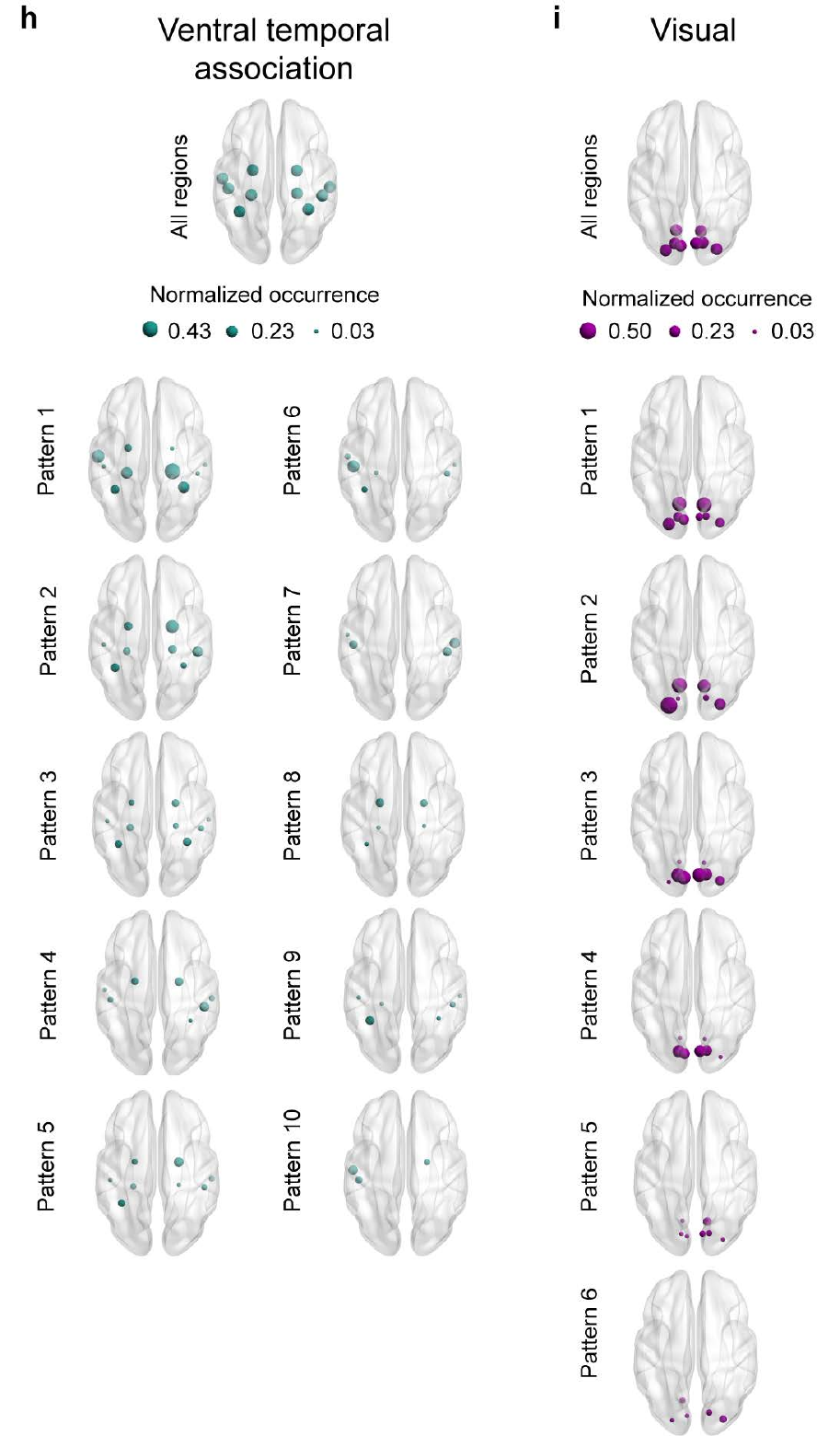}

\caption {Normalized contribution of brain regions to the prevalent patterns of synchronization. In Fig. 5 of the main text we show the prevalent patterns of synchronization that we observed after systematic regional stimulation within the brain across 30 individuals. Here, we show the brain regions (nodes) that contributed in producing these patterns within each cognitive system. Each sub-figure represents a cognitive system and the first image depicts all the brain regions within the system. In the following images, we show only the brain regions (nodes) that produced the corresponding patterns in Fig. 5. The size of the node represents the normalized occurrence for the given pattern. In panel (b), in order to help the reader differentiate superior and transverse regions, we mark them with S and T respectively. From this figure it can be clearly observed that different nodes within the same cognitive system can produce different patterns. Normalized occurrences for different patterns also vary across patterns and across cognitive systems.}
\label{fig:s63}
\end{figure}

\newpage
\begin{figure} 
\centering
\includegraphics [width=0.4\linewidth, keepaspectratio]
{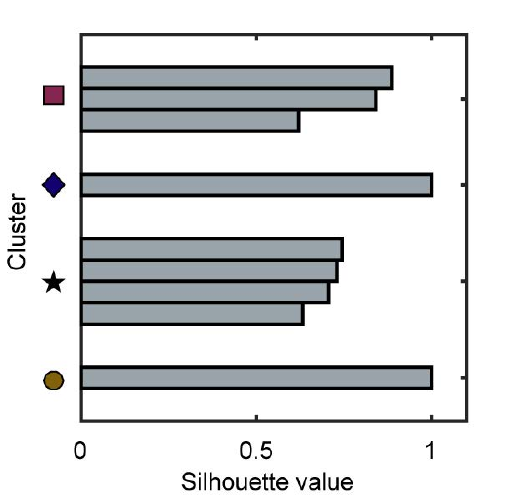}
\caption{Clustering of cognitive systems using patterns tenacity. As described in the main text, in the parameter space of subject and node tenacity we grouped cognitive systems into clusters using the k-means algorithm and silhouette analysis. For $k=4$ we observed an optimized clustering (Fig. 6). Here we show the corresponding silhouette plot using squared Euclidean distance of the data points from the centroid of the cluster to which they were assigned.}
%{\justifying{\textbf{Supplementary Figure 7: Clustering of cognitive systems using patterns tenacity.} As described in the main text, in the parameter space of subject and node tenacity we grouped cognitive systems into clusters using the k-means algorithm and silhouette analysis. For $k=4$ we observed an optimized clustering (Fig. 6). Here we show the corresponding silhouette plot using squared Euclidean distance of the data points from the centroid of the cluster to which they were assigned.}\par}
\label{fig:s7}
\end{figure}

\newpage
\FloatBarrier
\begin{table}[!htbp]
\centering
\begin{tabular}{|c|c|c|c|c|c|}
\hline
S.N.	& Name of the brain region	& Cognitive assignment	&S.N.	 &Name of the brain region	&Cognitive assignment
\\ \hline
1	&Lateral orbitofrontal	&Att	&20	&Paracentral	 &MS \\ \hline
2	&Superior parietal	&Att	&21	&Postcentral 	&MS\\ \hline
3	&Superior temporal	&Aud	&22	&Precentral	&MS \\ \hline
4	&Transverse temporal	&Aud	&23	&Thalamus	&SC \\ \hline
5	&Caudal anterior cingulate	 &CP	&24	&Caudate	&SC \\ \hline
6	&Pars opercularis	&CP	&25	&Putamen	&SC \\ \hline
7	&Pars orbitalis	&CP	&26	&Pallidum	&SC \\ \hline
8	&Rostral anterior cingulate	&CP	&27	&Hippocampus	&SC \\ \hline
9	&Rostral middle frontal	&CP	&28	&Amygdala	&SC \\ \hline
10	&Supramarginal	&CP	&29	&Accumbens	 &SC \\ \hline
11	&Caudal middle frontal	&FP	&30	&Entorhinal	&VT \\ \hline
12	&Inferior parietal	&FP	&31	&Fusiform	&VT \\ \hline
13	&Medial orbitofrontal	&FP	&32	&Inferior temporal	&VT \\ \hline
14	&Pars triangularis	&FP	&33	&Middle temporal	&VT \\ \hline
15	&Insula	&FP	&34	&Parahippocampal	&VT \\ \hline
16	&Isthmus cingulate	&mDM	&35	&Cuneus	&V \\ \hline
17	&Posterior cingulate	&mDM	&36	&Lateral occipital	&V \\ \hline
18	&Precuneus	&mDM	&37	&Lingual	&V \\ \hline
19	&Superior frontal	&mDM	&38	&Pericalcarine	&V \\ \hline

\end{tabular}
\caption {Assignment of brain regions to cognitive systems. Each node in the brain network is assigned to a predefined cognitive system. System assignments are the same for regions in both left and right hemisphere. In Figure S3, we show the distribution of regional brain volume for each cognitive system.}
\label{table:corr_post_tms}
\end{table}

\end{document}